\documentclass[pra,aps,floatfix,twocolumn,superscriptaddress,amsmath,amssymb]{revtex4-2}
\usepackage{graphicx}
\usepackage{xcolor}
\usepackage{bm}
\usepackage{hyperref}
\usepackage{mathtools}
\hypersetup{colorlinks=true,citecolor=blue,linkcolor=blue,urlcolor=blue}
\usepackage[T1]{fontenc} 
\usepackage{newtxtext,newtxmath}
\allowdisplaybreaks

\begin{document}

\title{%
Quantum many-body scars in the Bose-Hubbard model with a three-body constraint
}

\author{Ryui Kaneko}
\email{ryuikaneko@aoni.waseda.jp}
\affiliation{%
Waseda Research Institute for Science and Engineering, Waseda University, Shinjuku, Tokyo 169-8555, Japan}
\affiliation{%
Department of Engineering and Applied Sciences, Sophia University, Chiyoda, Tokyo 102-8554, Japan}
\affiliation{%
Department of Physics, Kindai University, Higashi-Osaka, Osaka 577-8502, Japan}

\author{Masaya Kunimi}
\email{kunimi@rs.tus.ac.jp}
\affiliation{%
Department of Physics, Tokyo University of Science, Shinjuku, Tokyo 162-8601, Japan}

\author{Ippei Danshita}
\email{danshita@phys.kindai.ac.jp}
\affiliation{%
Department of Physics, Kindai University, Higashi-Osaka, Osaka 577-8502, Japan}

\date{\today}

\begin{abstract}
We uncover the exact athermal eigenstates
in the Bose-Hubbard (BH) model with a three-body constraint,
motivated by the exact construction of
quantum many-body scar (QMBS) states in the $S=1$ $XY$ model.
These states are generated by applying
an $\rm SU(2)$ ladder operator consisting of
a linear combination of two-particle annihilation operators
to the fully occupied state.
By using the improved Holstein-Primakoff expansion,
we clarify that the QMBS states in the $S=1$ $XY$ model
are equivalent to those in the constrained BH model
with additional correlated hopping terms.
We also find that,
in the strong-coupling limit of the constrained BH model,
the QMBS state exists as
the lowest-energy eigenstate of the effective model
in the highest-energy sector.
This fact enables us to prepare the QMBS states
in a certain adiabatic process
and opens up the possibility of
observing them in ultracold-atom experiments.
\end{abstract}

\maketitle

\emph{Introduction.} 
Recent technological developments in
ultracold atoms in optical lattices~\cite{trotzky2012},
Rydberg atoms in optical-tweezer arrays~\cite{browaeys2020},
trapped-ion systems~\cite{blatt2012},
and 
superconducting qubit systems~\cite{neill2016}
allow for simulating dynamics in isolated quantum many-body systems
and enable us to observe thermalization in sufficiently large systems
in experiments.
One of the important concepts that partly explains
how isolated quantum many-body systems thermalize
is the strong eigenstate thermalization hypothesis
(strong ETH)~\cite{deutsch1991,srednicki1994,rigol2008}.
It claims that,
for all 
eigenstates of the quantum many-body Hamiltonian,
the expectation value of a local operator
coincides with that of the microcanonical ensemble,
and would cause the system to thermalize
after a long-time evolution~\cite{mori2018,dalessio2016}.
The strong ETH is often fulfilled
in nonintegrable systems without the extensive number
of conserved quantities~\cite{beugeling2014,kim2014},
but does not necessarily hold for
general nonintegrable systems~\cite{shiraishi2017}.
Indeed, such an ETH-breaking state has been observed in experiments
on nonintegrable systems prepared by Rydberg atoms
trapped in optical-tweezer
arrays~\cite{bernien2017,bluvstein2021}.

The discovery of the ETH-breaking state has stimulated
further studies on unconventional phenomena such as
the many-body localization~\cite{nandkishore2015,
choi2016,yan2017a,yan2017b,alet2018,abanin2019,mokhtarijazi2023},
the Hilbert-space fragmentation~\cite{sala2020,
khemani2020,yang2020,moudgalya2022b,moudgalya2022},
and the quantum many-body scar (QMBS)
states~\cite{turner2018,turner2018b,
james2019,shibata2020,mark2020,kuno2020,
papic2022,serbyn2021,moudgalya2022,yoshida2022,
chandran2023,sanada2023}.
Among others, the QMBS states,
in which the thermalization is extremely slow
or does not occur in isolated quantum many-body systems,
have gained significant attention
because of their observation
in a wide range of different models such as
the quantum Ising and PXP models
related to Rydberg-atom
systems~\cite{turner2018,turner2018b,
james2019,shiraishi2019,bull2019,lin2019,choi2019,ho2019,
mukherjee2020,lin2020,lin2020b,iadecola2020,michailidis2020,
sugiura2021,yao2022,kunimi2023_arxiv}
and optical-lattice systems~\cite{su2023}.
Several theoretical studies have recently addressed the QMBS states in
the Bose-Hubbard (BH) systems, which are commonly prepared with ultracold
atoms in optical lattices, including those in the classical limit
characterized by a high-dimensional chaotic phase space~\cite{hummel2023}
and those emerging due to the effects of correlated
hoppings~\cite{zhao2020,hudomal2020}.
However, their experimental observation is still lacking.

Although an emergent $\rm SU(2)$ algebra
that is not part of the symmetry group of the Hamiltonian
helps construct the QMBS state
and provides an intuitive understanding of its origin~\cite{choi2019},
to the best of our knowledge,
the emergent $\rm SU(2)$ algebra in the BH systems
has not been established yet.
Moreover,
although the $S=1$ $XY$ model is commonly treated
as a model for the strong-coupling limit
of the BH model~\cite{altman2002,huber2007,nagao2016},
the connection between the QMBS states~\cite{schecter2019,chattopadhyay2020}
[as well as the hidden $\rm SU(2)$ algebra~\cite{kitazawa2003}]
in the spin model
and 
those in the bosonic one
has not been thoroughly discussed.
If one can systematically construct the QMBS states
of BH systems
in a manner similar to other spin systems,
it would be much more helpful for
future ultracold-atom experiments.

In this Letter,
we construct the exact QMBS states
in the BH model with a three-body constraint.
To clarify the correspondence between
the $S=1$ $XY$ model and the constrained BH model,
we transform the spin model into the bosonic one
using the improved Holstein-Primakoff
expansion~\cite{lindgard1974,batyev1985,giacomo2016,vogl2020,koenig2021}
and find emergent correlated hopping terms,
which also possess the same QMBS states.
Furthermore, by considering the strong-coupling limit
of the constrained BH model,
we find that the QMBS state
corresponds to the lowest-energy eigenstate
of the effective model
in the highest-energy sector.
Based on this observation,
we discuss how to prepare and observe the QMBS state
in ultracold-atom systems.

\emph{Scars in the constrained BH model.} 
We consider the BH chain,
which is defined as
\begin{align}
\label{eq:hubbard_model_h_nmax_inf}
 \hat{H}^{\infty}
 &= \hat{H}^{\infty}_0
  + \hat{H}^{\infty}_{\rm int},
\\
 \hat{H}^{\infty}_0 &=
 - J \sum_{i} ( \hat{b}^{\dagger}_i \hat{b}_{i+1}
 + \mathrm{H.c.}),
 \quad\!
 \hat{H}^{\infty}_{\rm int} = 
   \frac{U}{2} \sum_{i} \hat{\nu}_i (\hat{\nu}_i - 1).
\end{align}
Here, the operators $\hat{b}_i$ and
$\hat{\nu}_i = \hat{b}^{\dagger}_i \hat{b}_i$ correspond to
the annihilation and particle number operators, respectively.
We take the lattice spacing to be unity
and focus on the even system size $L$.
The strengths of the hopping and interaction are represented as
$J$ and $U$, respectively.
The interaction $U$ can be both attractive and repulsive.
The superscript
$\infty$
indicates
that there is no restriction on the maximum occupation number.
We mainly choose open boundary conditions
in Eq.~\eqref{eq:hubbard_model_h_nmax_inf}
for numerical calculations
although the choice of boundary conditions does not affect
the presence of the QMBS states~\footnote{See the Supplemental Material.}

Hereafter,
we focus on the model with the maximum
occupation number $n_{\rm max} = 2$
(the occupation number at any site $i$
is restricted to be $n_i=0$, $1$, and $2$).
To this end, we apply the projection
$\hat{P}_{n_{\rm max} = 2}$ on each Hamiltonian
and obtain
\begin{align}
\label{eq:hubbard_model_h}
 \hat{H}
 &=
 \hat{P}_{n_{\rm max} = 2} \hat{H}^{\infty} \hat{P}_{n_{\rm max} = 2}
 =
 \hat{H}_0 + \hat{H}_{\rm int},
\\
 \hat{H}_0 &=
 - J \sum_{i} ( \hat{a}^{\dagger}_i \hat{a}_{i+1}
 + \mathrm{H.c.}),
 \quad
 \hat{H}_{\rm int} = 
   \frac{U}{2} \sum_{i} \hat{n}_i (\hat{n}_i - 1).
\end{align}
Here, the operators
$\hat{a}_i = \hat{P}_{n_{\rm max} = 2} \hat{b}_i \hat{P}_{n_{\rm max} = 2}$
and
$\hat{n}_i = \hat{a}^{\dagger}_i \hat{a}_i$ correspond to
the annihilation and particle number operators
after the projection, respectively.
When $J\not=0$,
the BH model (for $U\not=0$)~\cite{kolovsky2004}
and that with the constraint $n_{\rm max} = 2$
(for any $U$)~\cite{Note1}
are nonintegrable in general.
The majority of eigenstates of these nonintegrable models
should satisfy the volume-law scaling of the entanglement entropy (EE),
according to the ETH.
In contrast to these conventional states,
we will demonstrate that
the constraint model
possesses the QMBS states for any interaction $U$.

Inspired by the previous studies
on the $S=1$ $XY$ model~\cite{schecter2019,chattopadhyay2020},
we consider the
ladder operators
\begin{align}
 \hat{J}^{+} = \sum_{i} \frac{(-1)^{r_i}}{\sqrt{2}} \hat{a}_i^2,
\quad\!\!
 \hat{J}^{-} = \left( \hat{J}^{+} \right)^{\dagger}
~\text{with}\quad\!\!
 \hat{a} \rightarrow
 \begin{pmatrix}
   0 & 1 & 0 \\
   0 & 0 & \sqrt{2} \\
   0 & 0 & 0
 \end{pmatrix}
\end{align}
for the maximum occupation number $n_{\rm max}=2$.
Here,
the matrix representation of the operator $\hat{a}$
is obtained in the local Hilbert space spanned by
$\{|0\rangle, |1\rangle, |2\rangle\}$,
and
$r_i$ is the distance from the leftmost site ($r_i=i$).
The operators satisfy $\hat{a}^3 = (\hat{a}^{\dagger})^3 = 0$,
while
$\hat{a}\hat{a}^{\dagger} \not= \hat{a}^{\dagger} \hat{a} + 1$
with a three-body constraint.
From these,  
we obtain the commutation relation
$
 [\hat{a}_i^2, (\hat{a}^{\dagger}_j)^2]
 =
 \left(2 - 2 \hat{a}^{\dagger}_i \hat{a}_i\right) \delta_{ij}.
$
Using this relation, we define the operator
\begin{align}
 \hat{J}^z
 &=
 \frac{1}{2} [\hat{J}^{+},\hat{J}^{-}]
 =
 \frac{1}{2} \sum_{i} \left(1 - \hat{a}^{\dagger}_i \hat{a}_i\right).
\end{align}
The operators $\hat{J}^{\pm}$ and $\hat{J}^z$
obey an $\rm SU(2)$ algebra
($
 [\hat{J}^z,\hat{J}^{\pm}] = \pm \hat{J}^{\pm}
$)
since
$[\hat{a}^{\dagger}_i \hat{a}_i,\hat{a}_j^2] = - 2 \hat{a}_i^2 \delta_{ij}$
and
$[\hat{a}^{\dagger}_i \hat{a}_i,(\hat{a}^{\dagger}_j)^2] = 2 (\hat{a}^{\dagger}_i)^2 \delta_{ij}$.
It is clear that $[\hat{J}^z,\hat{H}]=0$
from $[\hat{J}^z,\hat{H}_0]=0$ and $[\hat{J}^z,\hat{H}_{\rm int}]=0$,
while $[\hat{J}^{\pm},\hat{H}]\not=0$ in general.
Note that the $\rm SU(2)$ algebra holds for $n_{\rm max} = 2$ by chance
and breaks down for $n_{\rm max}>2$~\cite{Note1}.

Using the properties of these ladder operators,
it is easy to show that the following states,
\begin{align}
\label{eq:def_bose_hubbard_scar}
 |S_n\rangle \propto (\hat{J}^+)^n |\Omega\rangle,
 \quad
 |\Omega\rangle = \bigotimes_{i}
 |2_i\rangle,
\end{align}
where $|2_i\rangle$ stands for $|n_i=2\rangle$,
correspond to the bosonic counterpart of
the QMBS states
found in the $S=1$ $XY$ model~\cite{schecter2019},
satisfying
$\hat{H}_0|S_n\rangle = 0$
and
$\hat{H}_{\rm int}|S_n\rangle \propto |S_n\rangle$.
It is given as
\begin{align}
\label{eq:def_sn}
 |S_n\rangle
 &=
 \sum_{i_1\not=\cdots\not=i_n}
 \frac{(-1)^{r_{i_1}+\cdots+r_{i_n}}}{\binom{L}{n}^{1/2}}
 \bigotimes_{j}
 \begin{cases}
 |0_j\rangle, & j\in \{i_1,\dots,i_n\}, \\
 |2_j\rangle, & \mathrm{otherwise},
 \end{cases}
\end{align}
for general $n$.
We will leave the detailed derivation for the Supplemental Material~\cite{Note1}
and discuss in which symmetry sector the QMBS states appear.
Under the space inversion symmetry operation
($\hat{\mathcal{I}}$),
the boson creation operators satisfy
$
 \hat{\mathcal{I}} \hat{a}^{\dagger}_i \hat{\mathcal{I}}
 = \hat{a}^{\dagger}_{L+1-i}.
$
The ladder operator fulfills
$
 \hat{\mathcal{I}} \hat{J}^{+} \hat{\mathcal{I}}
  = - \hat{J}^{+},
$
which results in
$
 \hat{\mathcal{I}} |S_n\rangle
 = (-1)^n |S_n\rangle.
$
This relation means that
the QMBS state has even (odd) parity for even (odd)
$n(=L-N/2)$ with $N$ being the total particle number.
Therefore, for even $L$,
we should focus on the sectors with
even parity $\mathcal{I}=+1$  
(odd parity $\mathcal{I}=-1$)
when $N=4m$ ($N=4m+2$)
with $m$ being an integer.

Because the 
QMBS
states of the BH model have exactly the same
structure as those in the $S=1$ $XY$ model,
they exhibit the same EE
and the equivalent energy.
The von Neumann
EE is defined as
$S^{\rm vN}_{A} = - \mathrm{Tr} \hat{\rho}_A \ln \hat{\rho}_A$
with $\hat{\rho}_A$ being the reduced density matrix
for a region $A$ of size $L_A$.
When $L_A=L/2$,
$S^{\rm vN}_{A}$
for the state $|S_{n=L/2}\rangle$
would be
$
 S^{\rm vN}_{A}(n=L/2) \rightarrow
 [\ln(\pi L/8)+1]/2
$
($L\rightarrow\infty$)~\cite{schecter2019}.
As for the energy,
because the state $|S_n\rangle$ is the eigenstate of both
$\hat{n}^{\rm tot} = \sum_{i} \hat{n}_i$ and
$\hat{d}^{\rm tot} = \sum_{i} \hat{n}_i^2$,
i.e.,
$
 \hat{n}^{\rm tot} |S_n\rangle
 =
 2(L-n) |S_n\rangle
$,
$
 \hat{d}^{\rm tot} |S_n\rangle
 =
 4(L-n) |S_n\rangle
$,
the equation
$
 \hat{H} |S_n\rangle
 =
 (\hat{H}_0 + \hat{H}_{\rm int}) |S_n\rangle
 =
 \frac{U}{2}N |S_n\rangle
$
holds with the number of total particles
$N=2(L-n)$ ($=0$, $2$, $4$, \dots, $2L-2$, $2L$).

\begin{figure}[t]
\centering
\includegraphics[width=\columnwidth]{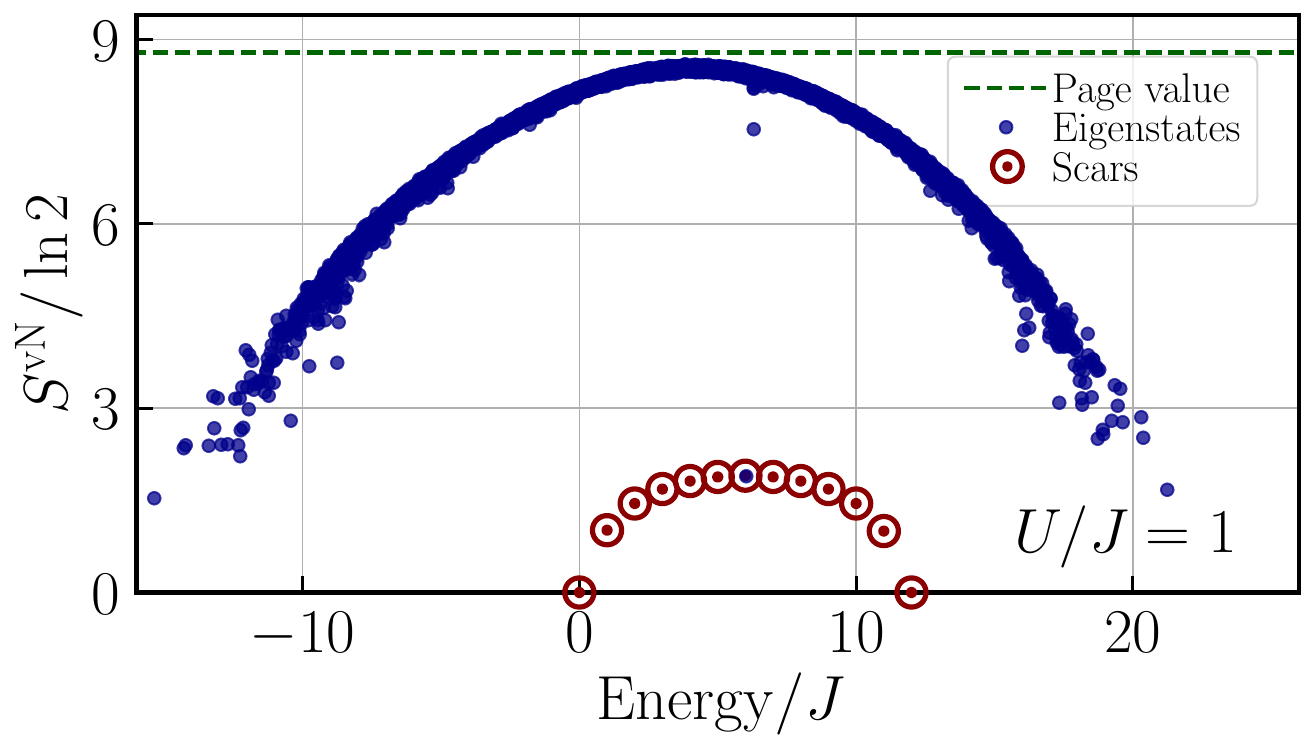}
\caption{%
EE as a function of energy.
The EE
of each state (a blue dot)
is given in the unit of $\ln 2$,
and its largest value almost saturates at the Page
value~\cite{page1993}
$S^{\rm Page} = (\ln 3)L/2 - 1/2$~\cite{schecter2019,chattopadhyay2020}
(a green dashed line).
We consider the system size $L=12$
under open boundary conditions
and the interaction strength 
$U/J=1$
in Eq.~\eqref{eq:hubbard_model_h}.
The quantum number sector with the particle number
$N=L$ (unit filling)
and the even parity $\mathcal{I}=+1$
is shown.
Each QMBS state (a red circle with a dot)
with the particle number $N$
has the energy $UN/2$
and the expectation values
$\langle\sum_{i=1}^{L} \hat{n}_i\rangle = N$ and
$\langle\sum_{i=1}^{L} \hat{n}_i^2\rangle = 2N$.
Its half-chain
EE at $N=L$, corresponding to a state $|S_{n=L/2}\rangle$,
becomes $S^{\rm vN}_{A}\rightarrow [\ln(\pi L/8)+1]/2$
for $L\rightarrow\infty$.}
\label{fig:scars}
\end{figure}

Numerical results on the corresponding
EE versus the energy
are presented in Fig.~\ref{fig:scars}.
Most of the eigenstates exhibit the volume-law scaling of EE,
whereas the QMBS states show the area-law scaling
(with a logarithmic correction) of EE.

\emph{Correspondence between the $S=1$ $XY$ model and the constrained BH model.} 
Let us first present the transformation that we use to study the
correspondence between the two models.
We utilize
the improved Holstein-Primakoff
expansion~\cite{lindgard1974,batyev1985,giacomo2016,vogl2020,koenig2021}
for $S=1$ spin operators
\begin{align}
 \hat{S}^{+}_i = \sqrt{2}\hat{b}_i + (1-\sqrt{2})\hat{b}^{\dagger}_i\hat{b}_i^2 +
 \left(\frac{1}{\sqrt{2}}-1\right)(\hat{b}^{\dagger}_i)^2\hat{b}_i^3,
\end{align}
with $\hat{b}_i$ being the boson annihilation operator
before the Hilbert space truncation.
The advantage over the conventional Holstein-Primakoff expansion is that
the Hilbert space on which the operator acts splits into
the physical ($n_i=0$, $1$, \dots, $n_{\rm max}=2$)
and unphysical  ($n_i=n_{\rm max}+1$, $n_{\rm max}+2$, \dots)
spaces~\cite{Note1}.
Therefore,
as long as the operator acts on the state in the physical subspace,
the generated states remain physical.
The transformation within the physical subspace
does not change the spectra of eigenenergies.
Consequently,
the bosonic operators
with the truncated Hilbert space $n_{\rm max}=2$
(namely,
$\hat{a}_i
 = \hat{P}_{n_{\rm max} = 2} \hat{b}_i \hat{P}_{n_{\rm max} = 2}$)
can be mapped exactly to $S=1$ spin operators.
Then, the spin ladder operator
$
 \hat{K}^{+} = \frac{1}{2} \sum_{i} (-1)^{r_i} (\hat{S}^{+}_i)^2,
$
which is used for constructing the QMBS states in the $S=1$ $XY$
model~\cite{schecter2019,chattopadhyay2020},
is evaluated as
\begin{align}
 \hat{K}^{+}
 &= \frac{1}{2} \sum_{i} (-1)^{r_i}
 \biggl[
 \sqrt{2} b_i^2
 - \sqrt{2} \hat{b}^{\dagger}_i \hat{b}_i^3
 + \frac{1}{\sqrt{2}} (\hat{b}^{\dagger}_i)^2 \hat{b}_i^4
\nonumber
\\
 &~\phantom{=}~
 + \left(5-3\sqrt{2}\right) (\hat{b}^{\dagger}_i)^3 \hat{b}_i^5
 + \left(\frac{3}{2}-\sqrt{2}\right) (\hat{b}^{\dagger}_i)^4 \hat{b}_i^6
 \biggr].
\end{align}
We immediately see $\hat{K}^{+} \rightarrow \hat{J}^{+}$
with a three-body constraint
($\hat{a}_i^3
 = \hat{P}_{n_{\rm max}=2} \hat{b}_i^3 \hat{P}_{n_{\rm max}=2}
 = 0$),
indicating the QMBS states in both systems are equivalent.

We then transform the term
$
 \hat{H}^{XY}_0
 =
 J_{xy} \sum_i ( \hat{S}^x_i \hat{S}^x_{i+1} + \hat{S}^y_i
\hat{S}^y_{i+1} )
$
in the $S=1$ $XY$ system
into the bosonic one.
Expanding it by the bosonic operator $\hat{b}_i$,
we get correlated hopping terms in addition to the conventional boson hopping term:
\begin{align}
\label{eq:s1_to_boson_corr_hop}
 \hat{H}^{XY}_0
 &=
 J_{xy} \sum_i
 (\hat{b}_i \hat{b}^{\dagger}_{i+1}
 + \mathrm{H.c.})
\nonumber
\\
 &~\phantom{=}~
 + \left( \frac{1}{\sqrt{2}}-1\right) J_{xy} \sum_i
 (
   \hat{b}_i \hat{b}^{\dagger}_{i+1} \hat{\nu}_{i+1}
 + \hat{b}_{i+1} \hat{b}^{\dagger}_i \hat{\nu}_i
 + \mathrm{H.c.}
 )
\nonumber
\\
 &~\phantom{==}~
 + \left(\frac{3}{2}-\sqrt{2}\right) J_{xy} \sum_i
 (
   \hat{\nu}_i \hat{b}_i \hat{b}^{\dagger}_{i+1} \hat{\nu}_{i+1}
  + \mathrm{H.c.}
 ).
\end{align}
Here, we drop unphysical higher-order terms, which
correspond to those containing $\hat{b}_i^3$
at the rightmost end.
Correlated hoppings are known to play a crucial role
in stabilizing the QMBS
states~\cite{zhao2020,hudomal2020,tamura2022}.
By utilizing the improved Holstein-Primakoff expansion,
we successfully show that
the correlated hopping terms in our model
also possess the same QMBS as in the original constrained
BH model~\cite{Note1}.

\emph{Scars in the strong-coupling limit.} 
Let us discuss how the QMBS states behave
in the strong-coupling limit,
which will be useful for experimental realization
as we will explain later.
We consider the strong-coupling limit of
the BH model on an open chain
with a three-body constraint at unit filling:
$
 \hat{H} =
 - J \sum_{i=1}^{L-1} ( \hat{a}^{\dagger}_i \hat{a}_{i+1}
 + \hat{a}^{\dagger}_{i+1} \hat{a}_i )
 + \sum_{i=1}^{L} \Omega_i \hat{n}_i
 + \frac{U}{2} \sum_{i=1}^{L} \hat{n}_i (\hat{n}_i - 1)
$.
Here, $\Omega_i$ is the external potential,
which is often chosen to be a parabolic one in experiments,
and the local Hilbert space is spanned by
$\{|0\rangle, |1\rangle, |2\rangle\}$.
We derive the effective model in the strong $U/J$ limit
for the Hilbert subspace satisfying
$\sum_i n_i = N$ (unit filling)
and
$n_i=0$ or $2$
using the Schrieffer-Wolff
transformation~\cite{cohentannoudji1998,bravyi2011}:
\begin{align}
\label{eq:effective_model}
 \hat{H}_{\mathrm{eff}} &=
   \frac{1}{2}LU
 + \sum_{i=1}^{L} \Omega_i
 + \frac{1}{2} \sum_{i=1}^{L-1} \tilde{J}_{i,i+1}
 + \sum_{i=1}^{L} \tilde{h}_i \hat{T}_i^z
\nonumber
\\
 &~\phantom{=}~
 + 2\sum_{i=1}^{L-1} \tilde{J}_{i,i+1}
 ( \hat{T}_i^x \hat{T}_{i+1}^x
 + \hat{T}_i^y \hat{T}_{i+1}^y
 - \hat{T}_i^z \hat{T}_{i+1}^z).
\end{align}
Here, we define
\begin{align}
 \tilde{h}_i &:= 2\Omega_i
 - (J_{i,i+1}^{+} - J_{i,i+1}^{-})
 + (J_{i-1,i}^{+} - J_{i-1,i}^{-}),
\\
 \tilde{J}_{i,i+1} &:= J_{i,i+1}^{+} + J_{i,i+1}^{-} 
 = \frac{2J^2U}{U^2 - (\Omega_{i+1}-\Omega_i)^2},
\\
 J_{i,i+1}^{\pm} &:=
 \begin{cases}
 \frac{J^2}{U\pm(\Omega_{i+1}-\Omega_{i})}, & i=1,2,\dots,L-1, \\
 0, & i=0,L.
 \end{cases}
\end{align}
The operators $\hat{T}^{\alpha}$ ($\alpha=x,y,z$) are
the $S=1/2$ spin operators that act on the space
spanned by $\{|0\rangle, |2\rangle\}$,
and satisfy
$\hat{T}^z = (|2\rangle\langle2|-|0\rangle\langle0|)/2$
and $\hat{T}^{+} = |2\rangle\langle0|$.
Note that a similar effective model was derived
previously~\cite{petrosyan2007,rosch2008,carleo2012,kunimi2021}
although they are different from the present one
which prohibits the hopping process containing $n_i>2$.

After performing a spin rotation around the $z$ axis
by $\pi$ radians for even sites,
the effective model
in Eq.~\eqref{eq:effective_model}
transforms into
the ferromagnetic Heisenberg model
in the absence of the external potential ($\Omega_i=0$).
The ground state after the transformation
is a trivial ferromagnetic state,
which includes a state
$
 |\psi_{\mathrm{GS}}^{\mathrm{FM}}\rangle
 \propto
 \hat{P}_{\mathrm{UF}}
 \bigotimes_{j=1}^{L} \frac{|0_j \rangle + |2_j \rangle}{\sqrt{2}}
$
at unit filling.
Here, $\hat{P}_{\mathrm{UF}}$
is a projection onto the space at unit filling  
($\sum_i n_i=L$).
Then, the corresponding ground state
of the original effective model
in Eq.~\eqref{eq:effective_model}
becomes
$
 |\psi_{\mathrm{GS}}^{\mathrm{AF}}\rangle
 \propto
 \hat{P}_{\mathrm{UF}}
 \bigotimes_{j=1}^{L}
 \frac{|0_j \rangle - (-1)^j |2_j \rangle}{\sqrt{2}}
$,
which is equivalent to $|S_{L/2}\rangle$
in Eq.~\eqref{eq:def_sn}.
Therefore, the exact ground state of the effective model
in the strong-coupling limit
is embedded in the spectra of eigenstates
of the constrained BH model.
Note that the effect of small external potential is
found to be negligible~\cite{Note1}.

\begin{figure}[t]
\centering
\includegraphics[width=\columnwidth]{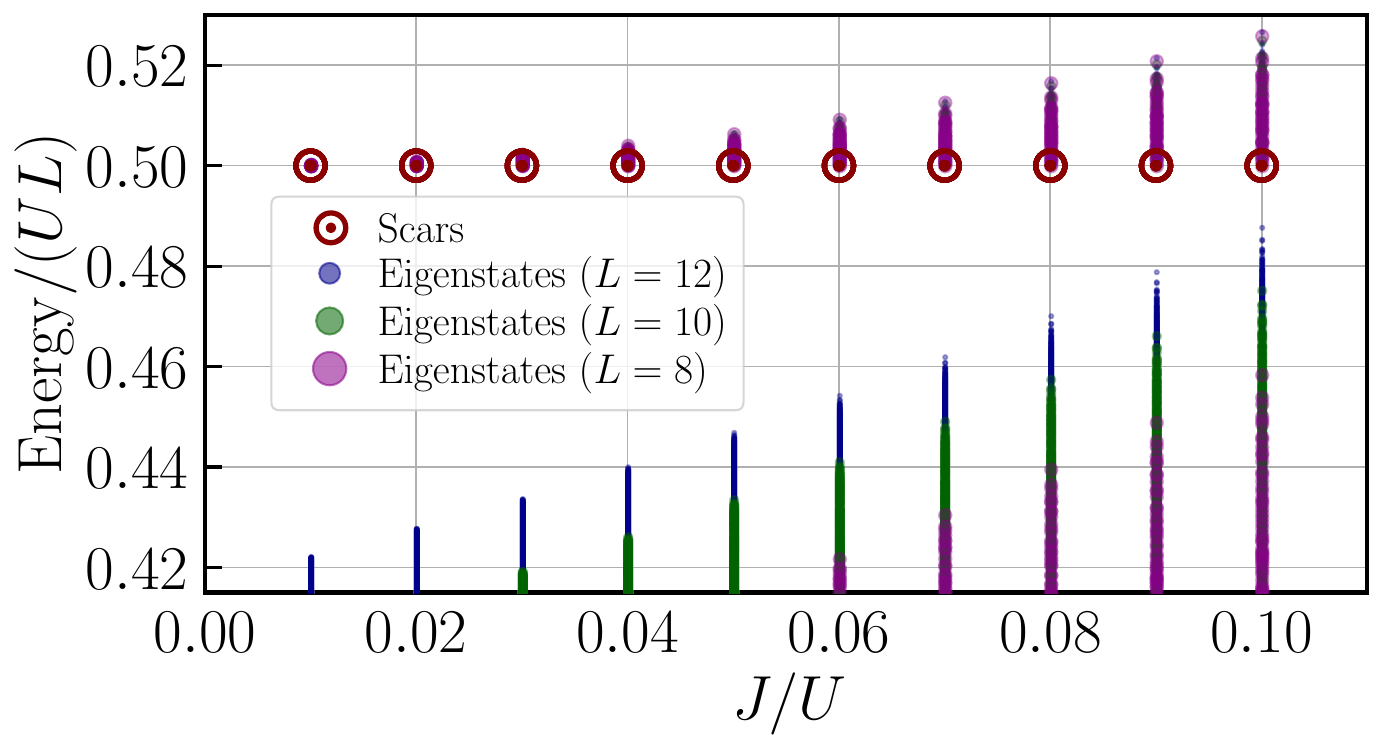}
\caption{Energy spectra as a function of
the interaction strength at unit filling.
We consider
the Hamiltonian in
Eq.~\eqref{eq:hubbard_model_h}
without the external potential
under open boundary conditions,
the system size $L=8$, $10$, and $12$,
and the interaction strength $J/U=0.01$,
$0.02$, $\dots$, $0.1$.
The quantum number sector with the particle number
$N=L$
and the parity $\mathcal{I}=(-1)^{{\rm mod}(L,4)/2}$
for $L=8$, $10$, and $12$
is given by largest purple, middle green, and smallest blue dots, respectively.
Each QMBS state (a red circle with a dot)
with the particle number $N(=L)$
has the energy $UN/2$.
The effective subspace exhibiting the highest energy ($\approx UL/2$)
and that exhibiting the second highest one ($\approx UL/2-U$)
are well separated energetically for $U\gg J$.
The QMBS state is found to become the lowest-energy eigenstate
in the former subspace for sufficiently strong interaction.
}
\label{fig:ene_spectra}
\end{figure}

Next, let us consider how the QMBS state behaves
in the case of $U/J<\infty$.
For $U\gg J$,
the energy spectrum is divided into well separated bands,
and the eigenstates belonging to each band
nearly preserve the number of sites
at which the particle number takes the values
$n_i=0$, $n_i=1$, and $n_i=2$.
In the case of unit filling with a three-body constraint,
the highest-energy band
(consisting of 
$L/2$ sites with $n_i=0$ and
$L/2$ sites with $n_i=2$ only)
and the second highest-energy band
(consisting of
$L/2-1$ sites with $n_i=0$,
$L/2-1$ sites with $n_i=2$, and
remaining $2$ sites with $n_i=1$)
are well separated energetically
(see Fig.~\ref{fig:ene_spectra}).
Because the QMBS state is the exact eigenstate
of the constrained BH model for any $U$,
we can always find the QMBS eigenstate
in the band consisting of only
$n_i=0$ and $n_i=2$ sites.
Remarkably, we have found that
the QMBS state corresponds to the lowest-energy eigenstate in this subspace
(the highest-energy band)
not only at the strong-coupling limit $J/U=0$
but also for $J/U\le 0.1$, using the exact diagonalization method
for $L\le 12$~\cite{Note1}.

\emph{Preparing scars in ultracold-atom experiments.} 
For a wide parameter region of $U\gg J$,
the QMBS state is found to become
the lowest-energy eigenstate
in the highest-energy band,
where the particle number takes only the values $0$ and $2$.
Therefore,
it would be possible to achieve QMBS states
in the BH model with a three-body constraint
in the following manner:
(i) Prepare the $202020\cdots$-type charge-density wave (CDW) state with
a choice of $\Omega_i = \mathrm{const} \times (-1)^{i}$
to join in the effective $n_i=0,2$ subspace~\cite{daley2009}.
(ii) Make the external potential nearly uniform 
($\Omega_i \approx \mathrm{const}$)
by adiabatically changing it
while keeping $U/J$ sufficiently strong.
The quantum adiabatic theorem ensures that
the final state approaches the lowest-energy eigenstate
(equivalent to $|S_{L/2}\rangle$) in the $n_i=0,2$ subspace.
(iii) Subsequently reduce the potential depth
so that the strength of the interaction becomes
comparable to the magnitude of hopping.
Note that
the ETH breaking may be caused by
the fragmentation~\cite{kunimi2021}
and the QMBS states for $U\gg J$.
To purely observe the effect of the QMBS states,
it is desirable to prepare the system with smaller $U$.

Concerning the three-body constraint,
strong three-body losses of atoms in optical lattices
prohibit more than two particles from occupying a single site
because of the continuous quantum Zeno effect,
resulting in the Bose-Hubbard model with
$n_{\rm max}=2$~\cite{daley2009,mark2012}.
To allow control over the ratio of
the three-body-loss term to the two-body interaction term,
one can use a broad Feshbach resonance~\cite{kraemer2006,tanzi2013,tanzi2016}.
This procedure enables us to realize
the much stronger three-body-loss term
than the interaction and hopping terms,
while keeping periodic potentials shallow
so that the interaction is not so strong.
Because our QMBS states
in the constrained BH model
can be realized
for both attractive and repulsive interactions,
they would be detected 
in ultracold atoms with three-body losses
in optical lattices.

In such QMBS states with a fixed particle number,
most of the physical quantities
(such as single-particle correlations~\cite{takasu2020},
density-density correlations~\cite{cheneau2012},
and the R\'{e}nyi
EE~\cite{islam2015,kaufman2016})
exhibit almost no time dependence,
which would be observed after a sudden quench in experiments.
The logarithmic size dependence of
the EE growth
would also provide a smoking gun for the existence of the QMBS states.

\emph{Conclusions.} 
Motivated by the exact construction of the QMBS states
in the $S=1$ $XY$ model, we have provided the equivalent athermal states
in the BH model with a three-body constraint.
To get insight into the mechanism of realizing the QMBS states
in the bosonic system,
we transform the spin model into the bosonic model
using the improved Holstein-Primakoff
transformation~\cite{lindgard1974,batyev1985,giacomo2016,vogl2020,koenig2021}
that does not mix the physical and unphysical Hilbert spaces
and consequently does not change the exact energy spectra.
The bosonic model obtained after the transformation
is similar to the conventional BH model,
but with the additional correlated hopping terms,
which also possess the same QMBS states.
Based on the fact that
the QMBS state corresponds to the lowest-energy eigenstate
of the effective model for the strong-coupling limit
with the highest-energy sector,
we propose the realization of the QMBS states
in ultracold-atom systems.
Moreover, such a local effective Hamiltonian,
which also possesses the QMBS states as its eigenstates,
would deepen our understanding of the scar phenomena.

Our findings will stimulate further research
on the QMBS states
in general BH models without any constraints,
which would be realized more easily in experiments
of ultracold atoms in optical lattices.
The construction of the QMBS states
in the general spin $S$ system
has been discussed recently~\cite{odea2020,tang2022},
and it could be extended to the BH model
by utilizing the improved Holstein-Primakoff
transformation~\cite{lindgard1974,batyev1985,giacomo2016,vogl2020,koenig2021}.
These QMBS states
do not have to be generated according to
the conventional $\rm SU(2)$ algebra~\cite{odea2020,tang2022}.
This topic will be left for a subject of future study.

\smallskip

\begin{acknowledgments}
The authors acknowledge fruitful discussions with
Shimpei Goto,
Daichi Kagamihara,
Hosho Katsura,
Mathias Mikkelsen,
and
Daisuke Yamamoto.
This work was financially supported by JSPS KAKENHI
(Grants
No.\ JP18H05228,
No.\ JP20K14389,
No.\ JP21H01014,
No.\ JP21K13855,
and
No.\ JP22H05268),
by MEXT Q-LEAP (Grant No.\ JPMXS0118069021),
and by JST FOREST (Grant No.\ JPMJFR202T).
The numerical computations were performed on computers at
the Supercomputer Center, the Institute for Solid State Physics,
the University of Tokyo.
\end{acknowledgments}

\onecolumngrid

\end{document}


\title{%
Supplemental Material:
Quantum many-body scars in the Bose-Hubbard model with a three-body constraint
}

\author{Ryui Kaneko}
\email{ryuikaneko@aoni.waseda.jp}
\affiliation{%
Waseda Research Institute for Science and Engineering, Waseda University, Shinjuku, Tokyo 169-8555, Japan}
\affiliation{%
Department of Engineering and Applied Sciences, Sophia University, Chiyoda, Tokyo 102-8554, Japan}
\affiliation{%
Department of Physics, Kindai University, Higashi-Osaka, Osaka 577-8502, Japan}

\author{Masaya Kunimi}
\email{kunimi@rs.tus.ac.jp}
\affiliation{%
Department of Physics, Tokyo University of Science, Shinjuku, Tokyo 162-8601, Japan}

\author{Ippei Danshita}
\email{danshita@phys.kindai.ac.jp}
\affiliation{%
Department of Physics, Kindai University, Higashi-Osaka, Osaka 577-8502, Japan}

\date{\today}

\maketitle

\tableofcontents

\section{Scars in the constrained Bose-Hubbard model}
\label{supp:sec:construction_exact}

\subsection{Primitive proof that the
quantum many-body scar
states are the eigenstates
of the constrained Bose-Hubbard model}
\label{supp:sec:athermal_eigenstates}

We will see that the following states
\begin{align}
\label{supp:eq:def_bose_hubbard_scar}
 \hat{J}^{+} = \sum_{i} \frac{(-1)^{r_i}}{\sqrt{2}}  \hat{a}_i^2,
 \quad\!\!\!
 |S_n\rangle \propto (\hat{J}^+)^n |\Omega\rangle,
 \quad\!\!\!
 |\Omega\rangle = \bigotimes_{i} |2_i\rangle
\end{align}
correspond to the bosonic counterpart of
athermal eigenstates found in the $S=1$ $XY$ model~\cite{schecter2019}.

To show that the state $|S_n\rangle$ is an eigenstate of
the Hamiltonian
\begin{align}
\label{supp:eq:hubbard_model_h}
 \hat{H}
 &=
 \hat{H}_0 + \hat{H}_{\rm int},
\\
\label{supp:eq:hubbard_model_h0_hint}
 \hat{H}_0 &=
 - J \sum_{i} ( \hat{a}^{\dagger}_i \hat{a}_{i+1}
 + \mathrm{H.c.}),
 \quad\!\!\!
 \hat{H}_{\rm int} =
   \frac{U}{2} \sum_{i} \hat{n}_i (\hat{n}_i - 1),
\end{align}
we work in the local basis characterized by the particle number,
consisting of
$|n_i\rangle = |0\rangle$, $|1\rangle$, and $|2\rangle$.
Then, it is easy to check
$\hat{H}_{\rm int} |S_n\rangle \propto |S_n\rangle$.
Both repulsive ($U>0$) and attractive ($U<0$) interactions
satisfy this relation.
Hereafter, we will check that they are also eigenstates of $\hat{H}_0$.

Let us first focus on the $n=1$ case, where the state is
\begin{align}
 |S_1\rangle
 &=
 \frac{1}{\sqrt{L}}(
 - |0222\cdots22\rangle
 + |2022\cdots22\rangle
\nonumber
\\
 &~\phantom{=}~
 - |2202\cdots22\rangle
 + \cdots
 +(-1)^{L} |22\cdots20\rangle
 )
\\
 &=
 \frac{1}{\sqrt{L}} \sum_{i} (-1)^{r_i}
 |\underbrace{22\cdots22}_{i-1}0\underbrace{22\cdots22}_{L-i}\rangle.
\end{align}
For simplicity, we consider periodic boundary conditions.
We will see later that same arguments hold for open boundary conditions.
Since the maximum occupation number is $n_{\rm max}=2$,
the noninteracting terms
$(\hat{a}^{\dagger}_{i-1} \hat{a}_i
 + \hat{a}^{\dagger}_i \hat{a}_{i-1} )$
and
$(\hat{a}^{\dagger}_i \hat{a}_{i+1}
 + \hat{a}^{\dagger}_{i+1} \hat{a}_i )$
act on the state $|S_1\rangle$
in the following manner:
\begin{align}
 &~\phantom{=}~
 (\hat{a}^{\dagger}_{i-1} \hat{a}_i
 + \hat{a}^{\dagger}_i \hat{a}_{i-1} )|S_1\rangle
\nonumber
\\
 &=
 (\hat{a}^{\dagger}_{i-1} \hat{a}_i
 + \hat{a}^{\dagger}_i \hat{a}_{i-1} )
 |\underbrace{22\cdots2}_{i-2}202\underbrace{2\cdots22}_{L-i-1}\rangle
\nonumber
\\
 &= 
 \sqrt{2}
 |\underbrace{22\cdots2}_{i-2}112\underbrace{2\cdots22}_{L-i-1}\rangle
\end{align}
and
\begin{align}
 &~\phantom{=}~
 (\hat{a}^{\dagger}_i \hat{a}_{i+1}
 + \hat{a}^{\dagger}_{i+1} \hat{a}_i )|S_1\rangle
\nonumber
\\
 &=
 (\hat{a}^{\dagger}_i \hat{a}_{i+1}
 + \hat{a}^{\dagger}_{i+1} \hat{a}_i )
 |\underbrace{22\cdots2}_{i-2}202\underbrace{2\cdots22}_{L-i-1}\rangle
\nonumber
\\
 &= 
 \sqrt{2}
 |\underbrace{22\cdots2}_{i-2}211\underbrace{2\cdots22}_{L-i-1}\rangle.
\end{align}
Here, any term represented as
$
 |\underbrace{\cdots}_{i-1}2\underbrace{\cdots}_{L-i}\rangle
$
in $|S_1\rangle$
vanishes
since $\hat{a}^{\dagger}_i |2_i\rangle = 0$.
Then,
\begin{align}
 \hat{H}_0 |S_1\rangle
 &
 \propto
 \frac{1}{\sqrt{L}} \sum_{i} (-1)^{r_i}
 |\underbrace{22\cdots2}_{i-2}112\underbrace{2\cdots22}_{L-i-1}\rangle
\nonumber
\\
 &~\phantom{=}~
 +
 \frac{1}{\sqrt{L}} \sum_{i} (-1)^{r_i}
 |\underbrace{22\cdots2}_{i-2}211\underbrace{2\cdots22}_{L-i-1}\rangle.
\end{align}
By shifting $i\rightarrow i-1$ in the second term
of the right-hand side, we have
\begin{align}
 \hat{H}_0 |S_1\rangle
 &\propto
 \frac{1}{\sqrt{L}} \sum_{i} (-1)^{r_i}
 |\underbrace{22\cdots2}_{i-2}112\underbrace{2\cdots22}_{L-i-1}\rangle
\nonumber
\\
 &~\phantom{=}~
 +
 \frac{1}{\sqrt{L}} \sum_{i} (-1)^{r_i-1}
 |\underbrace{22\cdots2}_{i-3}211\underbrace{2\cdots22}_{L-i}\rangle
\\
 &=
 \frac{1}{\sqrt{L}} \sum_{i} (-1)^{r_i}
 |\underbrace{22\cdots2}_{i-2}112\underbrace{2\cdots22}_{L-i-1}\rangle
\nonumber
\\
 &~\phantom{=}~
 -
 \frac{1}{\sqrt{L}} \sum_{i} (-1)^{r_i}
 |\underbrace{22\cdots2}_{i-2}112\underbrace{2\cdots22}_{L-i-1}\rangle
 =0.
\end{align}
Therefore, $|S_1\rangle$ is a zero-energy eigenstate of $\hat{H}_0$.

For $n=2$, the state is given as
\begin{align}
 |S_2\rangle
 &=
 \frac{1}{\sqrt{L(L-1)/2}}(
 - |002222\cdots222\rangle
\nonumber
\\
 &~\phantom{=}~
 + |020222\cdots222\rangle
 - |022022\cdots222\rangle
 + \cdots
\nonumber
\\
 &~\phantom{==}~
 - |200222\cdots222\rangle
 + |202022\cdots222\rangle
 + \cdots
\nonumber
\\
 &~\phantom{===}~
 +(-1)^{2L-1} |22\cdots200\rangle
 )
\\
 &=
 \frac{1}{\sqrt{L(L-1)/2}} \sum_{i}
 \biggl[
 -
 |\underbrace{22\cdots22}_{i-1}00\underbrace{22\cdots22}_{L-i-1}\rangle
\nonumber
\\
 &~\phantom{=}~
 +
 \sum_{j\not=1}
 (-1)^{r_j}
 |\underbrace{22\cdots22}_{i-1}0\underbrace{22\cdots22}_{j-1}0\underbrace{22\cdots22}_{L-i-j}\rangle
 \biggr].
\end{align}
Let us discuss how part of the noninteracting term acts on
each term of $|S_2\rangle$.
For example,
since $\hat{a}|0\rangle=\hat{a}^{\dagger}|2\rangle=0$,
it is clear that
\begin{align}
 (\hat{a}^{\dagger}_i \hat{a}_{i+1}
 + \hat{a}^{\dagger}_{i+1} \hat{a}_i )
 |\underbrace{\cdots}_{i-1}00\underbrace{\cdots}_{L-i-1}\rangle
 &=0,
\\
 (\hat{a}^{\dagger}_i \hat{a}_{i+1}
 + \hat{a}^{\dagger}_{i+1} \hat{a}_i )
 |\underbrace{\cdots}_{i-1}22\underbrace{\cdots}_{L-i-1}\rangle
 &=0.
\end{align}
Because $|S_2\rangle$ is antisymmetrized,
the sign of the state
$|\underbrace{\cdots}_{i-1}02\underbrace{\cdots}_{L-i-1}\rangle$
is opposite to that of the state
$|\underbrace{\cdots}_{i-1}20\underbrace{\cdots}_{L-i-1}\rangle$.
By acting the noninteracting term to the linear combination of these two states, we obtain 
\begin{align}
 &~\phantom{=}~
 (\hat{a}^{\dagger}_i \hat{a}_{i+1}
 + \hat{a}^{\dagger}_{i+1} \hat{a}_i )
 (
 |\underbrace{\cdots}_{i-1}02\underbrace{\cdots}_{L-i-1}\rangle
 -
 |\underbrace{\cdots}_{i-1}20\underbrace{\cdots}_{L-i-1}\rangle
 )
\nonumber
\\
 &=
 \sqrt{2}
 (
 |\underbrace{\cdots}_{i-1}11\underbrace{\cdots}_{L-i-1}\rangle
 -
 |\underbrace{\cdots}_{i-1}11\underbrace{\cdots}_{L-i-1}\rangle
 )
 =0.
\end{align}
Therefore, $(\hat{a}^{\dagger}_i \hat{a}_{i+1}
 + \hat{a}^{\dagger}_{i+1} \hat{a}_i ) |S_2\rangle = 0$
holds for any site $i$.
Thus, the state $|S_2\rangle$ is the zero-energy eigenstate of $\hat{H}_0$.

For general $n$, the state is given as
\begin{align}
\label{eq:def_sn}
 |S_n\rangle
 &=
 \sum_{i_1\not=\cdots\not=i_n}
 \frac{(-1)^{r_{i_1}+\cdots+r_{i_n}}}{\binom{L}{n}^{1/2}}
 \bigotimes_{j}
 \begin{cases}
 |0_j\rangle, & j\in \{i_1,\dots,i_n\}, \\
 |2_j\rangle, & \mathrm{otherwise}.
 \end{cases}
\end{align}
Repeating the same procedure as in the case of $n=2$,
it is clear that 
the state $|S_n\rangle$ is the zero-energy eigenstate of $\hat{H}_0$.
Note that
the mechanism of the quantum many-body scar
(QMBS) states can also be understood
by a restricted spectrum generating algebra
(RSGA)~\cite{moudgaly2020}
(see Sec.~\ref{supp:sec:appendix_rsga} for more details).

The athermal behavior of
the eigenstate $|S_{n=L/2}\rangle$ is closely related
to the nonergodic dynamics after a quantum quench
starting from the charge-density-wave (CDW)
initial state $|\cdots2020\cdots\rangle$
to a strongly interacting regime
in the 1D Bose-Hubbard model at unit filling~\cite{kunimi2021}.
The state $|S_{L/2}\rangle$
is a linear combination of CDW states,
consisting of $L/2$ $|0\rangle$'s and $L/2$ $|2\rangle$'s.
The maximum particle number per site effectively becomes $n_{\rm max}=2$
in the strong-coupling limit,
and this CDW initial state has a
small but nonzero overlap
$\binom{L}{L/2}^{-1/2}
\approx
\sqrt{\pi L/2} \times 2^{-L}$
with the athermal state $|S_{n=L/2}\rangle$.

\subsection{Effects of boundary conditions and further-neighbor hoppings}

As was discussed in 
the previous studies~\cite{schecter2019,chattopadhyay2020},
because the same arguments hold
without the $(\hat{a}^{\dagger}_L \hat{a}_1
 + \hat{a}^{\dagger}_1 \hat{a}_L )$ term,
what we have shown here is valid for open boundary conditions as well.
Adding further-neighbor hoppings at odd distances apart
$J_3$, $J_5$, $J_7$, \dots 
and adding a chemical potential term
do not change the arguments either.
These states can be athermal eigenstates even when the hopping has randomness.

\subsection{Comment on the algebra
of corresponding operators for the Bose-Hubbard model
without any constraints}

When the local Hilbert space is not truncated ($n_{\rm max}\rightarrow\infty$),
relevant operators fulfill an $\rm SU(1,1)$
algebra~\cite{gerry1991a,gerry1991b}
for which the corresponding operators satisfy
$[\hat{L}^{+}, \hat{L}^{-}] = - 2 \hat{L}^{z}$
[the sign on the right-hand side is negative
incontrast to the $\rm SU(2)$ algebra]
and
$[\hat{L}^{z}, \hat{L}^{\pm}] = \pm \hat{L}^{\pm}$.
In this case, we define the ladder operators
\begin{align}
 \hat{L}^{+} = \frac{1}{2} \sum_{i} (-1)^{r_i} (\hat{b}^{\dagger}_i)^2,
\quad
 \hat{L}^{-} = \frac{1}{2} \sum_{i} (-1)^{r_i} \hat{b}_i^2,
\end{align}
where the role of
creation and annihilation operators is reversed
and the prefactor changes from $1/{\sqrt{2}}$ to $1/2$
compared to the $\rm SU(2)$ case.
Using the commutation relation
$[\hat{b}_i^2,(\hat{b}^{\dagger}_j)^2]
= (4\hat{b}^{\dagger}_i \hat{b}_i + 2) \delta_{ij}$,
we obtain
\begin{align}
 \hat{L}^z
 =
 - \frac{1}{2} [\hat{L}^{+},\hat{L}^{-}]
 =
 \frac{1}{2}
 \sum_{i} \left(\hat{b}^{\dagger}_i \hat{b}_i + \frac{1}{2}\right).
\end{align}
It is easy to show
the commutation relation
$[\hat{L}^z,\hat{L}^{\pm}] = \pm \hat{L}^{\pm}$
using the relations
$[\hat{b}^{\dagger}_i \hat{b}_i,\hat{b}_j^2] = - 2 \hat{b}_i^2 \delta_{ij}$
and
$[\hat{b}^{\dagger}_i \hat{b}_i,(\hat{b}^{\dagger}_j)^2] = 2 (\hat{b}^{\dagger}_i)^2 \delta_{ij}$.

\section{Correspondence between
the spin and constrained Bose-Hubbard models}
\label{supp:sec:correspondence_spin}

We have constructed the exact QMBS states
in the Bose-Hubbard model with a three-body constraint
in the previous section.
To clarify the correspondence between
the QMBS states
in the Bose-Hubbard and $S=1$ spin models,
we transform the spin model into the Bose-Hubbard-like model exactly.
According to the improved Holstein-Primakoff
expansion~\cite{lindgard1974,batyev1985,giacomo2016,vogl2020,koenig2021},
the matrix representation of the $S=1$ spin operator
$\hat{S}^{+} = \hat{S}^x + i \hat{S}^y$
is obtained in the local Hilbert space spanned by
the unconstrained boson basis
$\{|0\rangle, |1\rangle, |2\rangle, |3\rangle, |4\rangle, |5\rangle, \dots\}$
as
\begin{align}
 \hat{S}^{+} &= \sqrt{2}\hat{b} + (1-\sqrt{2})\hat{b}^{\dagger}\hat{b}^2 +
 \left(\frac{1}{\sqrt{2}}-1\right)(\hat{b}^{\dagger})^2\hat{b}^3
\\
 &\rightarrow
\begin{pmatrix}
\textcolor{black}{0} & \textcolor{black}{\sqrt{2}} & \textcolor{black}{0} & \textcolor{lightgray}{0} & \textcolor{lightgray}{0} & \textcolor{lightgray}{0} & \cdots\\
\textcolor{black}{0} & \textcolor{black}{0} & \textcolor{black}{\sqrt{2}} & \textcolor{lightgray}{0} & \textcolor{lightgray}{0} & \textcolor{lightgray}{0} & \cdots\\
\textcolor{black}{0} & \textcolor{black}{0} & \textcolor{black}{0} & \textcolor{lightgray}{0} & \textcolor{lightgray}{0} & \textcolor{lightgray}{0} & \cdots\\
\textcolor{lightgray}{0} & \textcolor{lightgray}{0} & \textcolor{lightgray}{0} & 0 & -6 + 2 \sqrt{2} & 0 & \cdots\\
\textcolor{lightgray}{0} & \textcolor{lightgray}{0} & \textcolor{lightgray}{0} & 0 & 0 & - 8 \sqrt{5} + 3 \sqrt{10} & \cdots\\
\textcolor{lightgray}{0} & \textcolor{lightgray}{0} & \textcolor{lightgray}{0} & 0 & 0 & 0 & \cdots\\
\vdots & \vdots & \vdots & \vdots & \vdots & \vdots & \ddots
\end{pmatrix}.
\end{align}
Note that the matrix elements that are colored gray,
connecting the physical and unphysical subspaces,
are all zero.
In the main text,
we find that
the transformed Bose-Hubbard-like model contains
additional terms, which
share the same QMBS states
defined in Eq.~\eqref{supp:eq:def_bose_hubbard_scar}.

\subsection{Generator of an \texorpdfstring{$\rm SU(2)$}{SU(2)} algebra
for the $S=1$ $XY$ model}

Let us first briefly review the QMBS states
in the $S=1$ $XY$ model
on a chain~\cite{schecter2019,chattopadhyay2020}.
They exist in the model given as
\begin{align}
\label{supp:eq:spin_model_h}
 \hat{H}^{XY} &= \hat{H}^{XY}_0 + \hat{H}^{XY}_{\rm int},
\\ 
\label{supp:eq:spin_model_h0}
 \hat{H}^{XY}_0
 &=
 J_{xy} \sum_i ( \hat{S}^x_i \hat{S}^x_{i+1} + \hat{S}^y_i
\hat{S}^y_{i+1} ),
\\
\label{supp:eq:spin_model_hint}
 \hat{H}^{XY}_{\rm int}
 &=
   H_z \sum_i \hat{S}^z_i
 + D_z \sum_i (\hat{S}^z_i)^2,
\end{align}
where $\hat{S}^{\alpha}$ ($\alpha=x$, $y$, $z$) denotes the $S=1$ spin
operator,
$J_{xy}$ is the strength of
the transverse spin exchange interaction,
$H_z$ is the strength of the magnetic field,
and $D_z$ is the strength of the single-ion anisotropy.
We choose open or periodic boundary conditions
in Eqs.~\eqref{supp:eq:spin_model_h0} and \eqref{supp:eq:spin_model_hint}.
The ladder operators
\begin{align}
 \hat{K}^{\pm} &= \frac{1}{2} \sum_{i} (-1)^{r_i} (\hat{S}^{\pm}_i)^2,
\\
 \hat{K}^{z} &= \frac{1}{2} \sum_{i} \hat{S}^{z}_i
\end{align}
with $\hat{S}^{\pm}_i = \hat{S}^x_{i} \pm i \hat{S}^y_i$
were introduced~\cite{kitazawa2003,schecter2019,chattopadhyay2020}
to study the $\rm SU(2)$ symmetry
characterized by
\begin{align}
 [\hat{K}^{+}, \hat{K}^{-}] = 2\hat{K}^{z},
 \quad
 [\hat{K}^{z}, \hat{K}^{\pm}] = \pm \hat{K}^{\pm}.
\end{align}
The related QMBS states are
generated as
\begin{align}
 |\tilde{S}_n\rangle &\propto (\hat{K}^+)^n |\tilde{\Omega}\rangle,
 \quad
 |\tilde{\Omega}\rangle = \bigotimes_{i}|m_i=-1\rangle,
\end{align}
which satisfy
$\hat{H}^{XY}_0|\tilde{S}_n\rangle = 0$
and
$\hat{H}^{XY}_{\rm int}|\tilde{S}_n\rangle \propto
|\tilde{S}_n\rangle$.
They exhibit a subextensive entanglement entropy (EE) growth
at most logarithmically with system size.

\subsection{Mapping the interaction Hamiltonian}
\label{supp:sec:mapping_hp_transformation}

Hereafter,
we focus on the truncated Hilbert space
$n_{\rm max}=2$ and replace all the bosonic operators
$\hat{b}_i$ with $\hat{a}_i$ for simplicity.
The remaining terms
in Eq.~\eqref{supp:eq:spin_model_hint}
can be obtained using the relation
\begin{align}
 \hat{S}^{z}_i = S - \hat{a}^{\dagger}_i \hat{a}_i
 = \frac{n_{\rm max}}{2} - \hat{a}^{\dagger}_i \hat{a}_i.
\end{align}
For the periodic boundary condition, 
they become
\begin{align}
 \hat{H}^{XY}_{\rm int}
 &=
   D_z \sum_i \hat{n}_i^2
 - \left( D_z n_{\rm max} + H_z
 \right) \sum_i \hat{n}_i
\nonumber
\\
 &~\phantom{=}~
 + \left( D_z \frac{n_{\rm max}^2}{4} + H_z \frac{n_{\rm max}}{2}
 \right) \sum_i 1.
\end{align}
Comparing it with the Bose-Hubbard model up to a constant,
we have
\begin{align}
\label{supp:eq:hubbard_model_xy_int}
 {\hat{H}_{\rm int}}'
 &=
   \frac{U}{2} \sum_i \hat{n}_i (\hat{n}_i-1)
 - \mu \sum_i \hat{n}_i
\\
 &=
   \frac{U}{2} \sum_i \hat{n}_i^2
 - \left(\frac{U}{2} + \mu\right) \sum_i \hat{n}_i
\end{align}
with
\begin{align}
 U &= 2D_z,
\\
 \mu &= D_z (n_{\rm max}-1) + H_z.
\end{align}

\subsection{Zero-energy eigenstates of correlated hopping}

Here, we will demonstrate that
each correlated hopping term
\begin{align}
\label{supp:eq:s1_to_boson_corr_hop}
 \hat{H}^{XY}_0
 &=
 J_{xy} \sum_i
 (\hat{a}_i \hat{a}^{\dagger}_{i+1}
 + \mathrm{H.c.})
\nonumber
\\
 &~\phantom{=}~
 + \left( \frac{1}{\sqrt{2}}-1\right) J_{xy} \sum_i
 (
   \hat{a}_i \hat{a}^{\dagger}_{i+1} \hat{n}_{i+1}
 + \hat{a}_{i+1} \hat{a}^{\dagger}_i \hat{n}_i
 + \mathrm{H.c.}
 )
\nonumber
\\
 &~\phantom{==}~
 + \left(\frac{3}{2}-\sqrt{2}\right) J_{xy} \sum_i
 (
   \hat{n}_i \hat{a}_i \hat{a}^{\dagger}_{i+1} \hat{n}_{i+1}
  + \mathrm{H.c.}
 ),
\end{align}
obtained in the main text,
possesses the QMBS states
defined in Eq.~\eqref{supp:eq:def_bose_hubbard_scar}.

We focus on
the generalized Bose-Hubbard Hamiltonian
\begin{align}
\label{supp:eq:generalized_bose_hubbard_model}
 \hat{H}^{\rm Hub}
 &= \hat{H}_0
 + \hat{H}_{\rm corr,1}
 + \hat{H}_{\rm corr,2}
 + {\hat{H}_{\rm int}}',
\end{align}
where the middle two terms on the right-hand side
are defined as
\begin{align}
\label{supp:eq:generalized_bose_hubbard_model_corr_1}
 \hat{H}_{\rm corr,1}
 &=
 - J_{\rm corr,1} \sum_i \biggl(
   \hat{a}_i \hat{a}^{\dagger}_{i+1} \hat{n}_{i+1}
 + \hat{a}_{i+1} \hat{a}^{\dagger}_i \hat{n}_i
 + \mathrm{H.c.}
 \biggr),
\\
\label{supp:eq:generalized_bose_hubbard_model_corr_2}
 \hat{H}_{\rm corr,2}
 &=
 - J_{\rm corr,2} \sum_i \left(
   \hat{n}_i \hat{a}_i \hat{a}^{\dagger}_{i+1} \hat{n}_{i+1}
  + \mathrm{H.c.}
 \right)
\end{align}
while 
$\hat{H}_0$ and
${\hat{H}_{\rm int}}'$ are defined in 
Eq.~\eqref{supp:eq:hubbard_model_h0_hint}
and
Eq.~\eqref{supp:eq:hubbard_model_xy_int}, respectively.
The original $S=1$ $XY$ model
[see Eqs.~\eqref{supp:eq:spin_model_h}--\eqref{supp:eq:spin_model_hint}]
is reproduced when we choose
\begin{align}
\label{supp:eq:spin_eq_hubbard_condition_1}
 J &= -J_{xy},
\\
\label{supp:eq:spin_eq_hubbard_condition_2}
 J_{\rm corr,1} &= - \left(\frac{1}{\sqrt{2}}-1\right) J_{xy},
\\
\label{supp:eq:spin_eq_hubbard_condition_3}
 J_{\rm corr,2} &= - \left(\frac{3}{2}-\sqrt{2}\right) J_{xy}.
\end{align}
On the other hand,
when the strengths of the correlated hoppings
$J_{\rm corr,1}$ and $J_{\rm corr,2}$
are chosen independently,
there will be no direct correspondence between
the $S=1$ $XY$ model 
and the Bose-Hubbard model.

As we have shown in Sec.~\ref{supp:sec:mapping_hp_transformation},
with a three-body constraint 
satisfying $n_{\rm max}=2$,
both the Bose-Hubbard model
and the $S=1$ $XY$ model share the equivalent athermal state
$|S_n\rangle$,
i.e.,
\begin{align}
 \hat{H}_{0}|S_n\rangle &= 0,
\\
 (\hat{H}_{0} + \hat{H}_{\rm corr,1}
 + \hat{H}_{\rm corr,2})|S_n\rangle &= 0,
\\
 {\hat{H}_{\rm int}}'|S_n\rangle &\propto |S_n\rangle
\end{align}
when
Eqs.~\eqref{supp:eq:spin_eq_hubbard_condition_1}--\eqref{supp:eq:spin_eq_hubbard_condition_3}
hold.
From this, the relation
\begin{align}
 ( \hat{H}_{\rm corr,1} + \hat{H}_{\rm corr,2} ) |S_n\rangle
 = 0
\end{align}
clearly holds.
We may further expect that
the state $|S_n\rangle$ is also the zero-energy eigenstate of
each of the correlated hopping terms
$\hat{H}_{\rm corr,1}$ and $\hat{H}_{\rm corr,2}$,
i.e.,
\begin{align}
 \hat{H}_{\rm corr,1}|S_n\rangle
 = \hat{H}_{\rm corr,2}|S_n\rangle
 = 0,
\end{align}
irrespective of the strength of the correlated hopping
$J_{\rm corr,1}$ or $J_{\rm corr,2}$.
This expectation is indeed true as we will prove
in Sec.~\ref{supp:sec:appendix_proof_corr}.
This result suggests that
the QMBS states exist in a wide range of parameter regions
for the Bose-Hubbard-like model,
as well as in the constrained Bose-Hubbard model
with a conventional hopping term
and in that with the correlated hopping terms corresponding to
the $S=1$ $XY$ model.
Note that other types of QMBS states
in the presence of these correlated hoppings were also studied
recently~\cite{zhao2020,hudomal2020},
separately from the interest in the $S=1$ $XY$ model.

\subsection{Primitive proof that the
quantum many-body scar
states are the eigenstates of the
correlated hopping terms}
\label{supp:sec:appendix_proof_corr}

We show that
the QMBS states of the constrained Bose-Hubbard model
with the conventional hopping term
are the eigenstates of the correlated hopping terms.
(As for the relation to the RSGA,
see also Sec.~\ref{supp:sec:appendix_rsga}.)

\subsubsection{In the case of
\texorpdfstring{$\hat{H}_{\rm corr,1}$}{Hcorr,1}}

Here, we will show that $|S_n\rangle$ is a zero-energy eigenstate of $\hat{H}_{\rm corr,1}$.

For $n=1$, each term in $\hat{H}_{\rm corr,1}$ acts on the state
$
 |S_1\rangle   
 =   
 \frac{1}{\sqrt{L}} \sum_{i} (-1)^{r_i}
 |\underbrace{22\cdots22}_{i-1}0\underbrace{22\cdots22}_{L-i}\rangle
$
in the following manner:
\begin{align}
 &~\phantom{=}~
 (\hat{a}_{i-1} \hat{a}^{\dagger}_i \hat{n}_i + \hat{n}_i \hat{a}_i \hat{a}^{\dagger}_{i-1})|S_1\rangle
\nonumber
\\
 &=
 (\hat{a}_{i-1} \hat{a}^{\dagger}_i \hat{n}_i + \hat{n}_i \hat{a}_i \hat{a}^{\dagger}_{i-1})
 |\underbrace{22\cdots2}_{i-2}202\underbrace{2\cdots22}_{L-i-1}\rangle
 =
 0,
\\
 &~\phantom{=}~
 (\hat{a}_i \hat{a}^{\dagger}_{i+1} \hat{n}_{i+1} + \hat{n}_{i+1} \hat{a}_{i+1} \hat{a}^{\dagger}_i)|S_1\rangle
\nonumber
\\
 &=
 (\hat{a}_i \hat{a}^{\dagger}_{i+1} \hat{n}_{i+1} + \hat{n}_{i+1} \hat{a}_{i+1} \hat{a}^{\dagger}_i)
 |\underbrace{22\cdots2}_{i-2}202\underbrace{2\cdots22}_{L-i-1}\rangle
\nonumber
\\
 &=
 \sqrt{2}
 |\underbrace{22\cdots2}_{i-2}211\underbrace{2\cdots22}_{L-i-1}\rangle,
\\
 &~\phantom{=}~
 (\hat{a}_i \hat{a}^{\dagger}_{i-1} \hat{n}_{i-1} + \hat{n}_{i-1} \hat{a}_{i-1} \hat{a}^{\dagger}_i)|S_1\rangle
\nonumber
\\
 &=
 (\hat{a}_i \hat{a}^{\dagger}_{i-1} \hat{n}_{i-1} + \hat{n}_{i-1} \hat{a}_{i-1} \hat{a}^{\dagger}_i)
 |\underbrace{22\cdots2}_{i-2}202\underbrace{2\cdots22}_{L-i-1}\rangle
\nonumber
\\
 &=
 \sqrt{2}
 |\underbrace{22\cdots2}_{i-2}112\underbrace{2\cdots22}_{L-i-1}\rangle,
\\
 &~\phantom{=}~
 (\hat{a}_{i+1} \hat{a}^{\dagger}_i \hat{n}_i + \hat{n}_i \hat{a}_i \hat{a}^{\dagger}_{i+1})|S_1\rangle
\nonumber
\\
 &=
 (\hat{a}_{i+1} \hat{a}^{\dagger}_i \hat{n}_i + \hat{n}_i \hat{a}_i \hat{a}^{\dagger}_{i+1})
 |\underbrace{22\cdots2}_{i-2}202\underbrace{2\cdots22}_{L-i-1}\rangle
 =
 0.
\end{align}
Here, any term represented as
$
 |\underbrace{\cdots}_{i-1}2\underbrace{\cdots}_{L-i}\rangle
$
in $|S_1\rangle$
vanishes
since $\hat{a}^{\dagger}_i |2_i\rangle = 0$.
Then,
\begin{align}
 \hat{H}_{\rm corr,1} |S_1\rangle
 &\propto
 \frac{1}{\sqrt{L}} \sum_{i} (-1)^{r_i}
 |\underbrace{22\cdots2}_{i-2}112\underbrace{2\cdots22}_{L-i-1}\rangle
\nonumber
\\
 &~\phantom{=}~
 +
 \frac{1}{\sqrt{L}} \sum_{i} (-1)^{r_i}
 |\underbrace{22\cdots2}_{i-2}211\underbrace{2\cdots22}_{L-i-1}\rangle.
\end{align}
As in the case of $\hat{H}_0$,
by shifting $i\rightarrow i-1$ in the second term
of the right-hand side, the first and second terms cancel each other out.
Thus, $|S_1\rangle$ is a zero-energy eigenstate of $\hat{H}_{\rm corr,1}$.

For $n=2$, as in the case of $\hat{H}_0$, we consider
how part of the noninteracting term acts on
each term of $|S_2\rangle$.
First, it is clear that
\begin{align}
 (\hat{a}_i \hat{a}^{\dagger}_{i+1} \hat{n}_{i+1} + \hat{n}_{i+1} \hat{a}_{i+1} \hat{a}^{\dagger}_i)
 |\underbrace{\cdots}_{i-1}00\underbrace{\cdots}_{L-i-1}\rangle
 &=0,
\\
 (\hat{a}_i \hat{a}^{\dagger}_{i+1} \hat{n}_{i+1} + \hat{n}_{i+1} \hat{a}_{i+1} \hat{a}^{\dagger}_i)
 |\underbrace{\cdots}_{i-1}22\underbrace{\cdots}_{L-i-1}\rangle
 &=0,
\\
 (\hat{a}_{i+1} \hat{a}^{\dagger}_i \hat{n}_i + \hat{n}_i \hat{a}_i \hat{a}^{\dagger}_{i+1})
 |\underbrace{\cdots}_{i-1}00\underbrace{\cdots}_{L-i-1}\rangle
 &=0,
\\
 (\hat{a}_{i+1} \hat{a}^{\dagger}_i \hat{n}_i + \hat{n}_i \hat{a}_i \hat{a}^{\dagger}_{i+1})
 |\underbrace{\cdots}_{i-1}22\underbrace{\cdots}_{L-i-1}\rangle
 &=0.
\end{align}
Then, we focus on terms
$|\underbrace{\cdots}_{i-1}02\underbrace{\cdots}_{L-i-1}\rangle$
and
$|\underbrace{\cdots}_{i-1}20\underbrace{\cdots}_{L-i-1}\rangle$,
whose coefficients have the opposite sign
due to the antisymmetrization of $|S_2\rangle$.
A straightforward calculation yields
\begin{align}
 &~\phantom{=}~
 (\hat{a}_i \hat{a}^{\dagger}_{i+1} \hat{n}_{i+1} + \hat{n}_{i+1} \hat{a}_{i+1} \hat{a}^{\dagger}_i)
\nonumber
\\
 &~\phantom{==}~
 \cdot
 (
 |\underbrace{\cdots}_{i-1}02\underbrace{\cdots}_{L-i-1}\rangle
 -
 |\underbrace{\cdots}_{i-1}20\underbrace{\cdots}_{L-i-1}\rangle
 )
\nonumber
\\
 &
 =
 \sqrt{2}
 |\underbrace{\cdots}_{i-1}11\underbrace{\cdots}_{L-i-1}\rangle,
\\
 &~\phantom{=}~
 (\hat{a}_{i+1} \hat{a}^{\dagger}_i \hat{n}_i + \hat{n}_i \hat{a}_i \hat{a}^{\dagger}_{i+1})
\nonumber
\\
 &~\phantom{==}~
 \cdot
 (
 |\underbrace{\cdots}_{i-1}02\underbrace{\cdots}_{L-i-1}\rangle
 -
 |\underbrace{\cdots}_{i-1}20\underbrace{\cdots}_{L-i-1}\rangle
 )
\nonumber
\\
 &
 =
 - \sqrt{2}
 |\underbrace{\cdots}_{i-1}11\underbrace{\cdots}_{L-i-1}\rangle,
\\
 \therefore
 &~\phantom{=}~
 (\hat{a}_i \hat{a}^{\dagger}_{i+1} \hat{n}_{i+1} + \hat{n}_{i+1} \hat{a}_{i+1} \hat{a}^{\dagger}_i
 + \hat{a}_{i+1} \hat{a}^{\dagger}_i \hat{n}_i + \hat{n}_i \hat{a}_i \hat{a}^{\dagger}_{i+1})
\nonumber
\\
 &~\phantom{==}~
 \cdot
 (
 |\underbrace{\cdots}_{i-1}02\underbrace{\cdots}_{L-i-1}\rangle
 -
 |\underbrace{\cdots}_{i-1}20\underbrace{\cdots}_{L-i-1}\rangle
 )
 = 0.
\end{align}
Therefore, $(\hat{a}_i \hat{a}^{\dagger}_{i+1} \hat{n}_{i+1} + \hat{n}_{i+1} \hat{a}_{i+1} \hat{a}^{\dagger}_i
 + \hat{a}_{i+1} \hat{a}^{\dagger}_i \hat{n}_i + \hat{n}_i \hat{a}_i \hat{a}^{\dagger}_{i+1}) |S_2\rangle = 0$
for any site $i$.
Thus, the state $|S_2\rangle$ is the zero-energy eigenstate of
$\hat{H}_{\rm corr,1}$.

Repeating the same procedure as in the case of $n=2$,
we can show that
the state $|S_n\rangle$ is the zero-energy eigenstate of $\hat{H}_{\rm corr,1}$
for $n>2$.

\subsubsection{In the case of
\texorpdfstring{$\hat{H}_{\rm corr,1}$}{Hcorr,2}}

Likewise, we can show that $|S_n\rangle$ is a zero-energy eigenstate of $\hat{H}_{\rm corr,2}$.

For $n=1$, each term in $\hat{H}_{\rm corr,2}$ acts on the state
$
 |S_1\rangle   
 =   
 \frac{1}{\sqrt{L}} \sum_{i} (-1)^{r_i}
 |\underbrace{22\cdots22}_{i-1}0\underbrace{22\cdots22}_{L-i}\rangle
$
in the following manner:
\begin{align}
 &~\phantom{=}~
 (\hat{n}_{i-1} \hat{a}_{i-1} \hat{a}^{\dagger}_i \hat{n}_i + \hat{n}_i \hat{a}_i \hat{a}^{\dagger}_{i-1} \hat{n}_{i-1})|S_1\rangle
\nonumber
\\
 &=
 (\hat{n}_{i-1} \hat{a}_{i-1} \hat{a}^{\dagger}_i \hat{n}_i + \hat{n}_i \hat{a}_i \hat{a}^{\dagger}_{i-1} \hat{n}_{i-1})
\nonumber
\\
 &~\phantom{==}~
 \cdot
 |\underbrace{22\cdots2}_{i-2}202\underbrace{2\cdots22}_{L-i-1}\rangle
\nonumber
\\
 &=
 0,
\end{align}
and
\begin{align}
 &~\phantom{=}~
 (\hat{n}_i \hat{a}_i \hat{a}^{\dagger}_{i+1} \hat{n}_{i+1} + \hat{n}_{i+1} \hat{a}_{i+1} \hat{a}^{\dagger}_i \hat{n}_i)|S_1\rangle
\nonumber
\\
 &=
 (\hat{n}_i \hat{a}_i \hat{a}^{\dagger}_{i+1} \hat{n}_{i+1} + \hat{n}_{i+1} \hat{a}_{i+1} \hat{a}^{\dagger}_i \hat{n}_i)
\nonumber
\\
 &~\phantom{==}~
 \cdot
 |\underbrace{22\cdots2}_{i-2}202\underbrace{2\cdots22}_{L-i-1}\rangle
\nonumber
\\
 &=
 0.
\end{align}
Here, any term represented as
$
 |\underbrace{\cdots}_{i-1}2\underbrace{\cdots}_{L-i}\rangle
$
in $|S_1\rangle$
vanishes
since $\hat{a}^{\dagger}_i |2_i\rangle = 0$.
Therefore,
acting any Hermitian term in $\hat{H}_{\rm corr,2}$ to $|S_1\rangle$ gives $0$,
and hence, 
$|S_1\rangle$ is a zero-energy eigenstate of $\hat{H}_{\rm corr,2}$.

For $n=2$, as in the cases of $\hat{H}_0$ and $\hat{H}_{\rm corr,1}$, we consider
how part of the noninteracting term acts on
each term of $|S_2\rangle$.
First, it is clear that
\begin{align}
 (\hat{n}_i \hat{a}_i \hat{a}^{\dagger}_{i+1} \hat{n}_{i+1} + \hat{n}_{i+1} \hat{a}_{i+1} \hat{a}^{\dagger}_i \hat{n}_i)
 |\underbrace{\cdots}_{i-1}00\underbrace{\cdots}_{L-i-1}\rangle
 &=0,
\\
 (\hat{n}_i \hat{a}_i \hat{a}^{\dagger}_{i+1} \hat{n}_{i+1} + \hat{n}_{i+1} \hat{a}_{i+1} \hat{a}^{\dagger}_i \hat{n}_i)
 |\underbrace{\cdots}_{i-1}22\underbrace{\cdots}_{L-i-1}\rangle
 &=0.
\end{align}
Then, we focus on terms
$|\underbrace{\cdots}_{i-1}02\underbrace{\cdots}_{L-i-1}\rangle$
and
$|\underbrace{\cdots}_{i-1}20\underbrace{\cdots}_{L-i-1}\rangle$,
whose coefficients have the opposite sign.
Again, a straightforward calculation yields
\begin{align}
 &~\phantom{=}~
 (\hat{n}_i \hat{a}_i \hat{a}^{\dagger}_{i+1} \hat{n}_{i+1} + \hat{n}_{i+1} \hat{a}_{i+1} \hat{a}^{\dagger}_i \hat{n}_i)
\nonumber
\\
 &~\phantom{==}~
 \cdot
 (
 |\underbrace{\cdots}_{i-1}02\underbrace{\cdots}_{L-i-1}\rangle
 -
 |\underbrace{\cdots}_{i-1}20\underbrace{\cdots}_{L-i-1}\rangle
 )
 = 0.
\end{align}
Therefore, $(\hat{n}_i \hat{a}_i \hat{a}^{\dagger}_{i+1} \hat{n}_{i+1} + \hat{n}_{i+1} \hat{a}_{i+1} \hat{a}^{\dagger}_i \hat{n}_i)
|S_2\rangle = 0$
for any site $i$.
Thus, the state $|S_2\rangle$ is the zero-energy eigenstate of
$\hat{H}_{\rm corr,2}$.

Repeating the same procedure as in the case of $n=2$,
$\hat{H}_{\rm corr,2} |S_n\rangle = 0$ holds for $n>2$.

\section{Analysis by the restricted spectrum generating algebra method}
\label{supp:sec:appendix_rsga}

The mechanism of the QMBS states is well understood
by an RSGA~\cite{moudgaly2020}.
In this section, we briefly review the concept of RSGA
and how it is satisfied in the case of our models.

\subsection{Restricted spectrum generating algebra of order one}

Let us consider the Hamiltonian $\hat{H}_0$,
its eigenstate $|\psi_0\rangle$,
and the operator $\hat{Q}^{\dagger}$
satisfying $\hat{Q}^{\dagger}|\psi_0\rangle\not=0$.
If the conditions
\begin{align}
 \label{supp:eq:condition1_RSGA1}
 \hat{H}_0|\psi_0\rangle &= E_0 |\psi_0\rangle,
\\
 \label{supp:eq:condition2_RSGA1}
 [\hat{H}_0, \hat{Q}^{\dagger}]|\psi_0\rangle &= \mathcal{E}\hat{Q}^{\dagger}|\psi_0\rangle,
\\
 \label{supp:eq:condition3_RSGA1}
 [[\hat{H}_0,\hat{Q}^{\dagger}],\hat{Q}^{\dagger}] &= 0
\end{align}
hold, the Hamiltonian $\hat{H}_0$ is said to exhibit
the restricted spectrum generating algebra of order one
(RSGA-1)~\cite{moudgaly2020}.

It is known that the eigenstates generated by the RSGA are the QMBS
states~\cite{moudgaly2020}.
In particular, for any state
\begin{align}
 |\psi_n\rangle := (\hat{Q}^{\dagger})^n|\psi_0\rangle (\not=0),
\end{align}
the relation
\begin{align}
 \label{supp:eq:results_RSGA}
 \hat{H}_0|\psi_n\rangle = (E_0+n\mathcal{E})|\psi_n\rangle
\end{align}
holds. In general, to confirm the presence of the QMBS states,
we have only to examine whether the Hamiltonian satisfies
the conditions in
Eqs.~\eqref{supp:eq:condition1_RSGA1}--\eqref{supp:eq:condition3_RSGA1}.

Hereafter, we will check how our models satisfy the RSGA-1.

\subsection{Constrained Bose-Hubbard model with a conventional hopping term}

We first consider the Hamiltonian
$\hat{H} = \hat{H}_0 + \hat{H}_{\rm int}$
given in Eq.~\eqref{supp:eq:hubbard_model_h}
under open or periodic boundary conditions.
When we choose
\begin{align}
 |\psi_0\rangle = |22\cdots 22\rangle,
\end{align}
it satisfies
\begin{align}
 \label{supp:eq:psi0_is_eigenstate_of_cBose_Hubbard}
 \hat{H}|\psi_0\rangle = LU|\psi_0\rangle.
\end{align}
We then calculate the commutators
$[\hat{H}_0, \hat{J}^+]$ and
$[\hat{H}_{\rm int}, \hat{J}^+]$.
For this purpose,
we write each operator as
\begin{align}
 \hat{a}_{j}
 &=
 |0_{j}\rangle\langle 1_{j}|
 +\sqrt{2}|1_{j}\rangle\langle 2_{j}|,
\\
 \hat{a}^{\dagger}_{j}
 &=
 |1_{j}\rangle\langle 0_{j}|
 +\sqrt{2}|2_{j}\rangle\langle 1_{j}|,
\\
 \hat{a}_{j}^2
 &=
 \sqrt{2}|0_{j}\rangle\langle 2_{j}|,
\\
 \hat{n}_{j}
 &=
 |1_{j}\rangle\langle 1_{j}|
 +2|2_{j}\rangle\langle 2_{j}|
\end{align}
and derive the commutation relations
\begin{align}
 [\hat{a}^{\dagger}_{j+1} \hat{a}_{j}, \hat{a}_{k}^2]
 &=
 (
 -2|0_{j+1}\rangle\langle 1_{j+1}|
\nonumber
\\
 &~\phantom{=}~
 +\sqrt{2}|1_{j+1}\rangle\langle 2_{j+1}|
 )
 \hat{a}_{j} \delta_{j+1,k},
\\
 [\hat{a}^{\dagger}_{j} \hat{a}_{j+1}, \hat{a}_{k}^2]
 &=
 (
 -2|0_{j}\rangle\langle 1_{j}|
 +\sqrt{2}|1_{j}\rangle\langle 2_{j}|
 )
 \hat{a}_{j+1} \delta_{jk},
\\
 [\hat{n}_{j},\hat{a}_{k}^2]
 &=
 - 2 \hat{a}_{j}^2 \delta_{jk},
\\
 [\hat{n}_{j}^2,\hat{a}_{k}^2]
 &=
 - 4 \hat{a}_{j}^2 \delta_{jk}.
\end{align}
By defining the two-site operator
\begin{align}
 \hat{h}_{ij}
 =
 |1_{i}0_{j}\rangle\langle 2_{i}1_{j}|
 -
 |0_{i}1_{j}\rangle\langle 1_{i}2_{j}|
\end{align}
and using the aforementioned relations,
we can simplify the commutator as
\begin{align}
 [\hat{H}_0, \hat{J}^+]
 &=
 - 3J \sum_{j} (-1)^{j} \hat{h}_{j,j+1},
\\
 [\hat{H}_{\rm int}, \hat{J}^+]
 &=
 - U \hat{J}^{+}.
\end{align}
Since $\hat{h}_{j,j+1}|\psi_0\rangle = 0$, we obtain
\begin{align}
 [\hat{H}_0, \hat{J}^+] |\psi_0\rangle = 0,
\end{align}
and consequently,
\begin{align}
 [\hat{H}, \hat{J}^+] |\psi_0\rangle = -U \hat{J}^{+} |\psi_0\rangle.
\end{align}
Finally, because the commutation relations
\begin{align}
 [\hat{h}_{j,j+1}, \hat{a}_{k}^2 \hat{1}_{k+1}] &= 0,
\\
 [\hat{h}_{j,j+1}, \hat{1}_{k} \hat{a}_{k+1}^2] &= 0
\end{align}
hold for any $j$ and $k$, we get
\begin{align}
 [[\hat{H},\hat{J}^{+}],\hat{J}^{+}] = 0.
\end{align}
Thus, the Hamiltonian $\hat{H}$ satisfies the RSGA-1.
The resulting QMBS states are represented as
\begin{align}
 |\psi_n\rangle &\propto (\hat{J}^{+})^{n} |22\cdots 22\rangle,
\\
 \hat{H} |\psi_n\rangle &= U(L-n) |\psi_n\rangle,
\end{align}
which are consistent with those obtained in the main text.

\subsection{With correlated hopping terms
\texorpdfstring{$\hat{H}_{\rm corr,1}$}{Hcorr1}
and
\texorpdfstring{$\hat{H}_{\rm corr,2}$}{Hcorr2}}

Next, we consider the Hamiltonian with the correlated hopping terms
$\hat{H}_{\rm corr,1}$ and
$\hat{H}_{\rm corr,2}$,
which are given in
Eq.~\eqref{supp:eq:generalized_bose_hubbard_model_corr_1}
and in Eq.~\eqref{supp:eq:generalized_bose_hubbard_model_corr_2},
respectively.

We can rewrite the Hamiltonians as
\begin{align}
 \hat{H}_{\rm corr,1}
 &=
 -J_{\rm corr,1} \sum_{j} \hat{h}^{\rm corr,1}_{j,j+1},
\\
 \hat{h}^{\rm corr,1}_{ij}
 &:=
   4 | 1_{i} 2_{j} \rangle \langle 2_{i} 1_{j} |
 + 4 | 2_{i} 1_{j} \rangle \langle 1_{i} 2_{j} |
\nonumber
\\
 &~\phantom{=}~
 + \sqrt{2} ( | 0_{i} 2_{j} \rangle + | 2_{i} 0_{j} \rangle ) \langle 1_{i} 1_{j} |
\nonumber
\\
 &~\phantom{=}\phantom{=}~
 + \sqrt{2} | 1_{i} 1_{j} \rangle ( \langle 0_{i} 2_{j} | + \langle 2_{i} 0_{j} | ) 
\end{align}
and
\begin{align}
 \hat{H}_{\rm corr,2}
 &=
 -J_{\rm corr,2} \sum_{j} \hat{h}^{\rm corr,2}_{j,j+1},
\\
 \hat{h}^{\rm corr,2}_{ij}
 &:=
   2 | 1_{i} 2_{j} \rangle \langle 2_{i} 1_{j} |
 + 2 | 2_{i} 1_{j} \rangle \langle 1_{i} 2_{j} |.
\end{align}
Using these representations,
we can show the relations
\begin{align}
 [\hat{h}^{\rm corr,1}_{j,j+1}, \hat{a}_{j}^2 \hat{1}_{j+1}] 
 &=
 2 | 1_{j} 1_{j+1} \rangle \langle 2_{j} 2_{j+1} |
\nonumber
\\
 &~\phantom{=}~
 - 2 | 0_{j} 0_{j+1} \rangle \langle 1_{j} 1_{j+1} |
\nonumber
\\
 &~\phantom{=}~
 - 4\sqrt{2} | 0_{j} 1_{j+1} \rangle \langle 1_{j} 2_{j+1} |,
\\
 [\hat{h}^{\rm corr,1}_{j,j+1}, \hat{1}_{j} \hat{a}_{j+1}^2 ] 
 &=
 2 | 1_{j} 1_{j+1} \rangle \langle 2_{j} 2_{j+1} |
\nonumber
\\
 &~\phantom{=}~
 - 2 | 0_{j} 0_{j+1} \rangle \langle 1_{j} 1_{j+1} |
\nonumber
\\
 &~\phantom{=}~
 - 4\sqrt{2} | 1_{j} 0_{j+1} \rangle \langle 2_{j} 1_{j+1} |
\end{align}
and
\begin{align}
 [\hat{h}^{\rm corr,2}_{j,j+1}, \hat{a}_{j}^2 \hat{1}_{j+1}] 
 &=
 -2\sqrt{2} | 0_{j} 1_{j+1} \rangle \langle 1_{j} 2_{j+1} |,
\\
 [\hat{h}^{\rm corr,2}_{j,j+1}, \hat{1}_{j} \hat{a}_{j+1}^2 ] 
 &=
 -2\sqrt{2} | 1_{j} 0_{j+1} \rangle \langle 2_{j} 1_{j+1} |.
\end{align}
After straightforward calculations, we obtain
\begin{align}
 [\hat{H}_{\rm corr,1}, \hat{J}^+]
 =
 - 4J_{\rm corr,1} \sum_{j} (-1)^{j} \hat{h}_{j,j+1},
\\
 [\hat{H}_{\rm corr,2}, \hat{J}^+]
 =
 - 2J_{\rm corr,2} \sum_{j} (-1)^{j} \hat{h}_{j,j+1}.
\end{align}
This result suggests that
\begin{align}
 [\hat{H}_0, \hat{J}^+]
 \propto
 [\hat{H}_{\rm corr,1}, \hat{J}^+]
 \propto
 [\hat{H}_{\rm corr,2}, \hat{J}^+].
\end{align}
As in the case of the Hamiltonian $\hat{H}_0$,
because both commutation relations
contain the term $\sum_{j} (-1)^{j} \hat{h}_{j,j+1}$
with $\hat{h}_{j,j+1} |\psi_0\rangle = 0$,
we have
\begin{align}
 [\hat{H}_{\rm corr,1}, \hat{J}^+] |\psi_0\rangle &= 0,
\\
 [\hat{H}_{\rm corr,2}, \hat{J}^+] |\psi_0\rangle &= 0.
\end{align}
It is also clear that
\begin{align}
 [[\hat{H}_{\rm corr,1},\hat{J}^{+}],\hat{J}^{+}] &= 0,
\\
 [[\hat{H}_{\rm corr,2},\hat{J}^{+}],\hat{J}^{+}] &= 0.
\end{align}
Thus, the Hamiltonian
$\hat{H} + \hat{H}_{\rm corr,1} + \hat{H}_{\rm corr,2}$
satisfies the RSGA-1.
The QMBS states for the Hamiltonian $\hat{H}$
are also those for the Hamiltonian
$\hat{H} + \hat{H}_{\rm corr,1} + \hat{H}_{\rm corr,2}$.

\subsection{Without the three-body constraint}

From the viewpoint of the RSGA,
we can understand how the present QMBS states break down
without the three-body constraint.
For soft-core bosons, we obtain
\begin{align}
 \hat{H} |22\cdots 22\rangle
 &=
 -J \sum_j
 (\hat{b}_{j+1}^{\dagger}\hat{b}_j+\hat{b}^{\dagger}_j\hat{b}_{j+1})
 |22\cdots 22\rangle
\nonumber
\\
 &~\phantom{=}~
 + LU
 |22\cdots 22\rangle
\\
 &=
 -\sqrt{6}J \sum_j
 (
 |\underbrace{22\cdots 22}_{j-1}13 \underbrace{22\cdots 22}_{L-j-1}\rangle
\nonumber
\\
 &~\phantom{=}~
 +|\underbrace{22\cdots 22}_{j-1}31 \underbrace{22\cdots 22}_{L-j-1}\rangle
 )
\nonumber
\\
 &~\phantom{=}\phantom{=}~
 + LU
 |22\cdots 22\rangle,
\end{align}
which means that the state $|\psi_0\rangle=|22\cdots 22\rangle$
is no longer the eigenstate of the Hamiltonian.
Therefore, the Hamiltonian does not satisfy the RSGA.
Note that,
although the present QMBS states do not survive
without the three-body constraint,
the existence of other types of QMBS states is not ruled out in general.

\section{Numerical results}
\label{supp:sec:numerical_results}

\subsection{Nonintegrability of each model}
\label{supp:sec:nonintegrability}

\begin{figure}[t]
\centering
\includegraphics[width=\columnwidth]{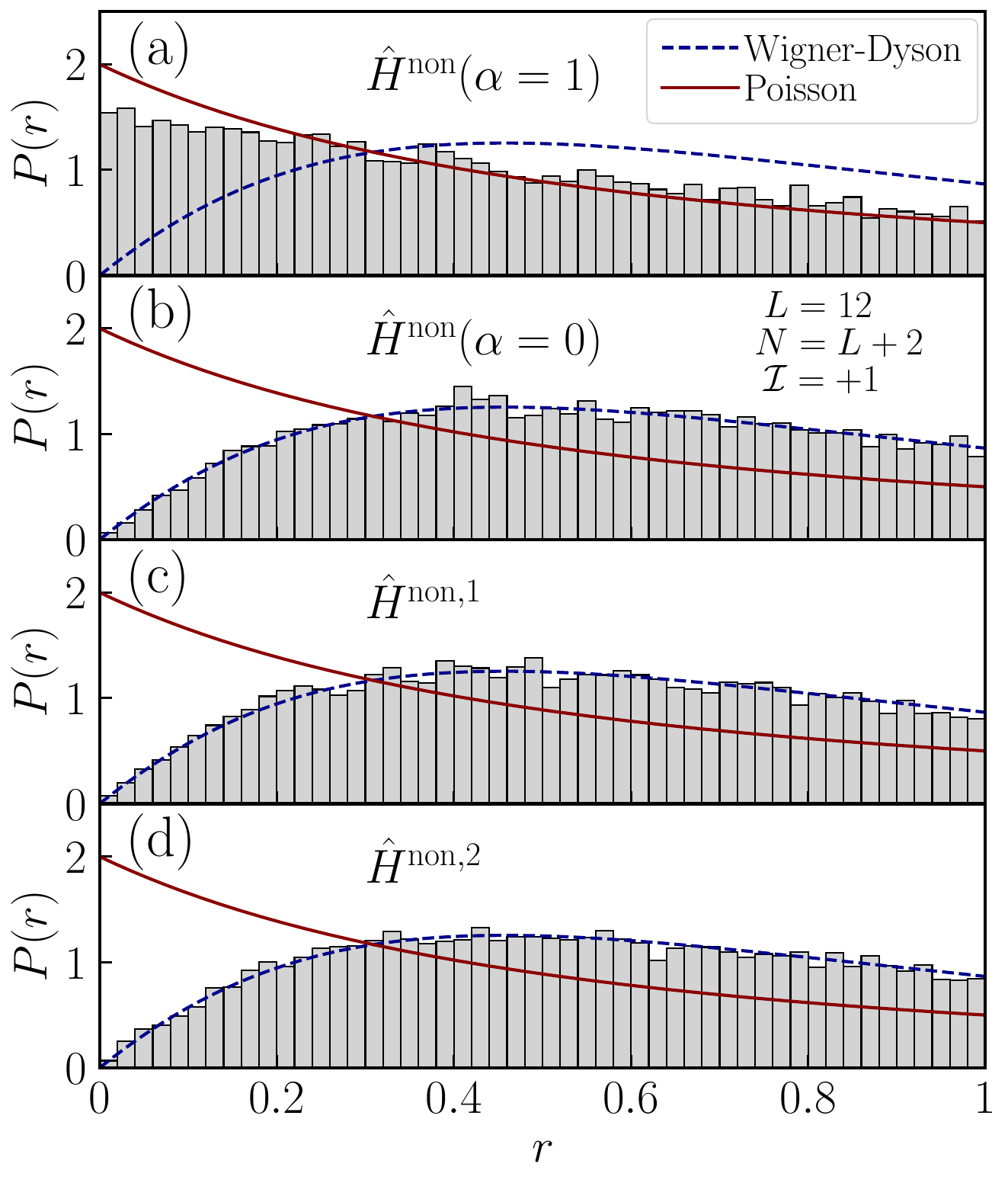}
\caption{%
Distribution $P(r)$ of the level spacing ratio.
The models
(a) $\hat{H}^{\rm non}(\alpha=1)$,
(b) $\hat{H}^{\rm non}(\alpha=0)$,
(c) $\hat{H}^{\rm non,1}$,
and
(d) $\hat{H}^{\rm non,2}$
for $L=12$
are chosen
[see Eqs.~\eqref{supp:eq:integrability_h_alpha}--\eqref{supp:eq:integrability_h_2}].
We focus on
the particle number $N=L+2$
in the even parity sector ($\mathcal{I}=+1$)
under open boundary conditions.
As for $\hat{H}^{\rm non}(\alpha=1)$,
$P(r)$ almost agrees with that of
the Poisson distribution,
suggesting that the model is integrable-like.
For the other models,
$P(r)$ exhibits the Wigner-Dyson distribution,
meaning that the model is nonintegrable.
}
\label{supp:fig:distribution_pr}
\end{figure}

\begin{figure}[t]
\centering
\includegraphics[width=\columnwidth]{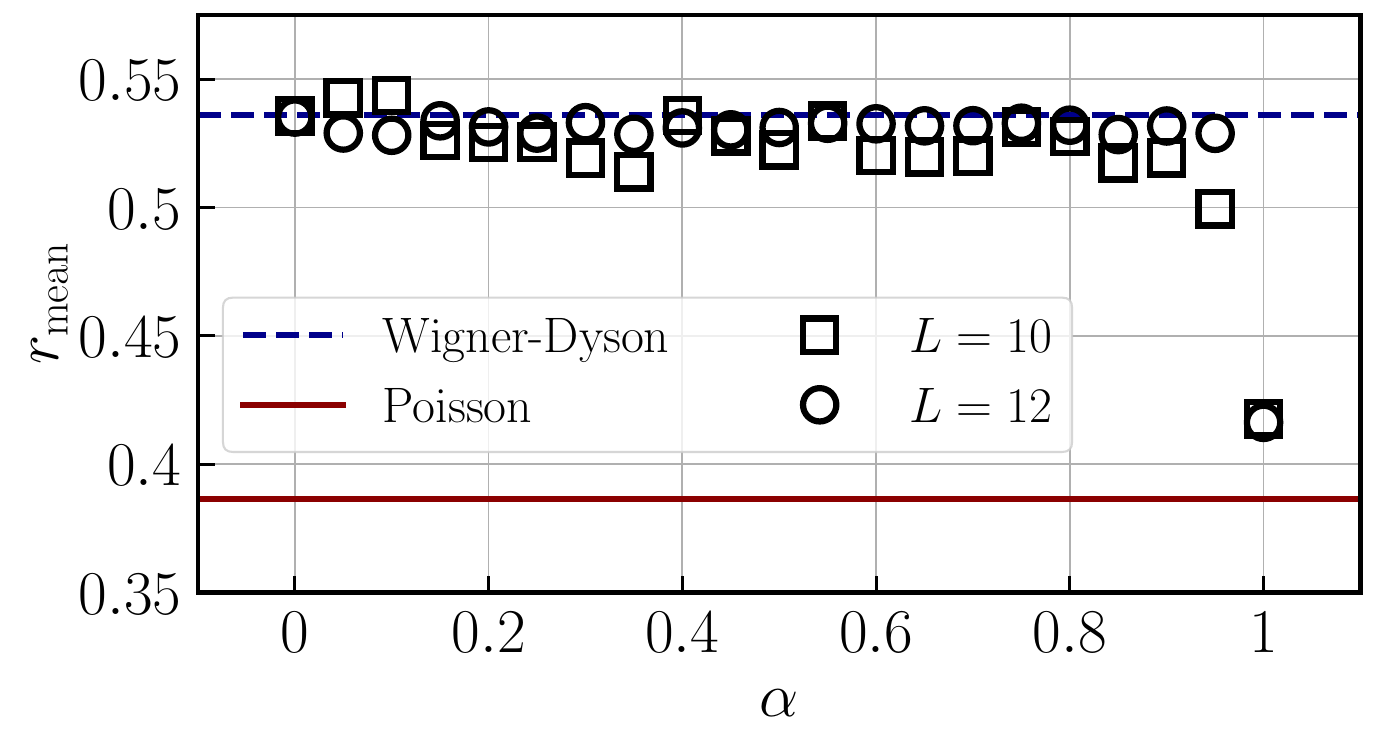}
\caption{%
Mean level spacing ratio $r_{\rm mean}$
as a function of $\alpha$
for the model $\hat{H}^{\rm non}(\alpha)$ defined in
Eq.~\eqref{supp:eq:integrability_h_alpha}.
Squares and circles correspond to those obtained for $L=10$
and $L=12$, respectively.
We consider the particle number $N=L+2$
in the even parity sector ($\mathcal{I}=+1$)
under open boundary conditions.
The model becomes the Bose-Hubbard model without any correlated hoppings
at $\alpha=0$, whereas it is equivalent to the $S=1$ $XY$ model at
$\alpha=1$.
For $\alpha<1$,
$r_{\rm mean}$ is close to the value
expected in the Wigner-Dyson distribution,
indicating that the model is nonintegrable in this parameter region.
}
\label{supp:fig:r_mean}
\end{figure}

\begin{figure}[!ht]
\centering
\includegraphics[width=\columnwidth]{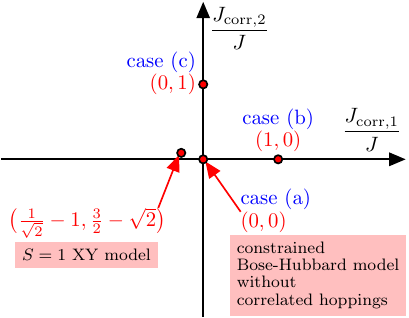}
\caption{%
Parameter points that we have chosen
to study the energy dependence of the EE.
See Fig.~\ref{supp:fig:scars_corr} for the numerical results.
}
\label{supp:fig:points}
\end{figure}

Before we confirm the presence of the QMBS states 
in the relevant Bose-Hubbard-like models numerically,
let us examine the nonintegrability of these models.
The Bose-Hubbard model for $U\not=0$ and $J\not=0$ is
nonintegrable~\cite{kolovsky2004},
and that at $U=0$ (the free boson model)
or $J=0$ (the atomic limit) is integrable.
On the other hand, 
additional constraints often work as effective interactions
even in the absence of explicit interaction terms.
It is less trivial whether the Bose-Hubbard model
with a three-body constraint or that with correlated hopping terms
is nonintegrable.
We will demonstrate that,
for $J\not=0$,
such models at $U=0$ are nonintegrable
in most cases,
whereas the model at $U=0$ accidentally becomes integrable-like
only when it is exactly mapped to the $S=1$ $XY$ model.

We focus on the following (noninteracting) Bose-Hubbard model
with the correlated hopping terms
under a three-body constraint
\begin{align}
\label{supp:eq:integrability_h_alpha}
 \hat{H}^{\rm non}(\alpha)
 &=
 \hat{H}_0
 +
 \hat{H}_{\rm corr,1}
 \left[J_{\rm corr,1}=\left(\frac{1}{\sqrt{2}}-1\right)\alpha J\right]
\nonumber
\\
 &~\phantom{=}~
 +
 \hat{H}_{\rm corr,2}
 \left[J_{\rm corr,2}=\left(\frac{3}{2}-\sqrt{2}\right)\alpha J\right],
\end{align}
where
$\hat{H}_0$, $\hat{H}_{\rm corr,1}$, and $\hat{H}_{\rm corr,2}$
are defined in
Eqs.~\eqref{supp:eq:hubbard_model_h0_hint},
\eqref{supp:eq:generalized_bose_hubbard_model_corr_1},
and
\eqref{supp:eq:generalized_bose_hubbard_model_corr_2}, respectively.
When $\alpha=0$, the model reduces to the Bose-Hubbard model
without the correlated hoppings.
On the other hand, when $\alpha=1$,
the model is equivalent to the $S=1$ $XY$ model.
The $S=1$ $XY$ model is known to exhibit
integrable-like behavior in the even total magnetization sector
because of a twisted $\rm SU(2)$ symmetry
in open and artificial boundary
conditions~\cite{kitazawa2003,schecter2019,chattopadhyay2020}.
(In the sector where the total spin is maximum,
the Hamiltonian becomes a constant matrix
and the system trivially exhibits integrable behavior.)
As we will see below,
we have examined the nonintegrability of these models
for $\alpha=0$, $0.05$, $0.1$, $\dots$, $0.95$, $1$
and found that the only $\alpha=1$ point is singular.

In addition,
we specifically consider the following models with
each correlated hopping term
\begin{align}
\label{supp:eq:integrability_h_1}
 \hat{H}^{\rm non,1}
 &=
 \hat{H}_0 + \hat{H}_{\rm corr,1}(J_{\rm corr,1}=J),
\\
\label{supp:eq:integrability_h_2}
 \hat{H}^{\rm non,2}
 &=
 \hat{H}_0 + \hat{H}_{\rm corr,2}(J_{\rm corr,2}=J).
\end{align}
As we will show below,
these models are found to be nonintegrable as well.
We will also demonstrate
how the QMBS states appear for these models
by looking into their
EE in the next section.

We investigate the nonintegrability of the models
by looking at the distribution $P(r)$ of the level spacing ratio,
which is defined as
\begin{align}
 r_n &= \frac{\min(s_n,s_{n+1})}{\max(s_n,s_{n+1})}
 \in [0,1],
\\
 s_n &= \frac{E_{n+1} - E_n}{J}
\end{align}
with $\{E_n\}$ being the ordered list of
eigenenergies~\cite{oganesyan2007,pal2010,atas2013,ho2018}. 
The model is inferred to be integrable-like (nonintegrable)
when $P(r)$ obeys the Poisson distribution $P(r)=2/(1+r)^2$
[the Wigner-Dyson distribution
$P(r) = (r+r^2)^{\beta}/(1+r+r^2)^{1+3\beta/2}/Z_{\beta}$
with $\beta$ being the Dyson index equal to $1$
and $Z_{\beta=1}=4/27$
for the Gaussian orthogonal ensemble]~\cite{ho2018}.
We also calculate
the mean level spacing ratio, which is given as
\begin{align}
 r_{\rm mean} = \left< r_n \right>.
\end{align}
Here, the symbol $\left< \cdots \right>$ represents
the average over all states in the given symmetric sector.
We compare the estimated mean level spacing ratio
with the expected $r_{\rm mean} = 4 - 2\sqrt{3} \approx 0.536$
(Wigner-Dyson distribution)
or $r_{\rm mean} = 2\ln 2 - 1 \approx 0.386$
(Poisson distribution)~\cite{atas2013}
to determine whether the model is likely to be nonintegrable.

\begin{figure}[!ht]
\centering
\includegraphics[width=\columnwidth]{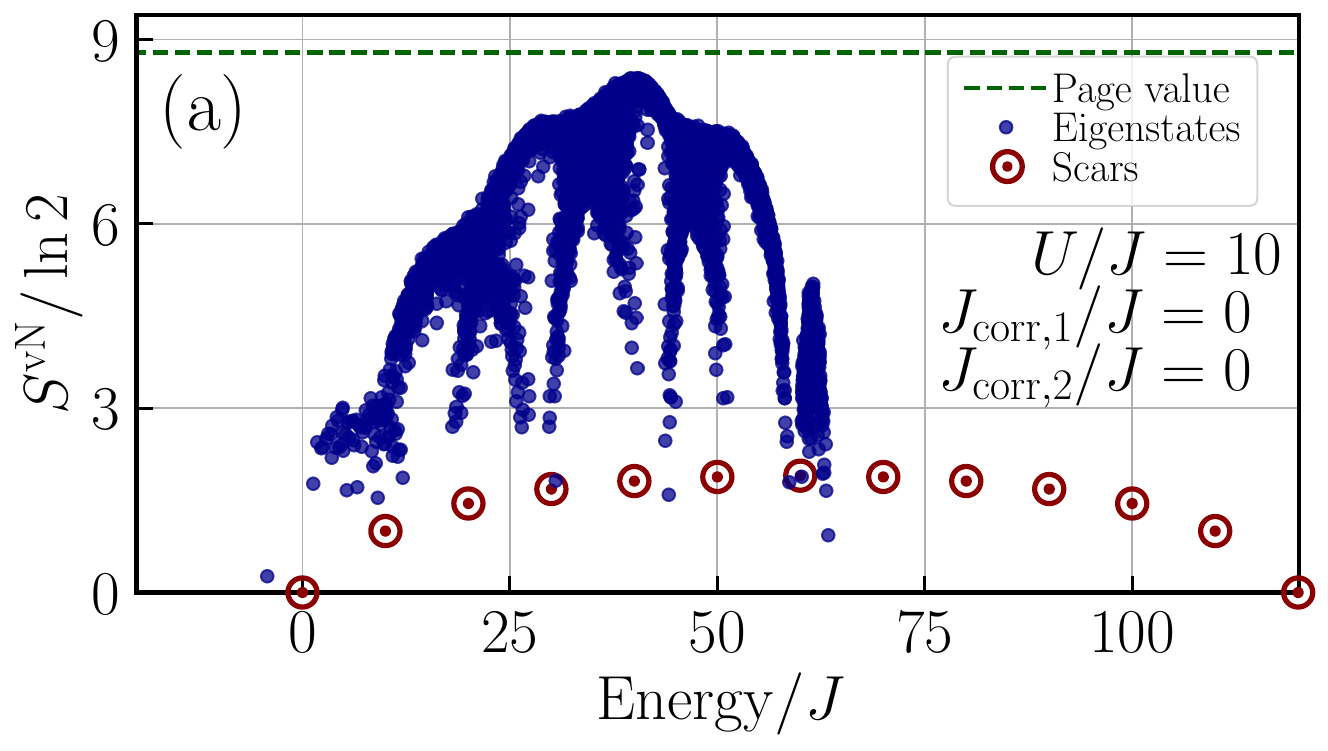}
\\
\includegraphics[width=\columnwidth]{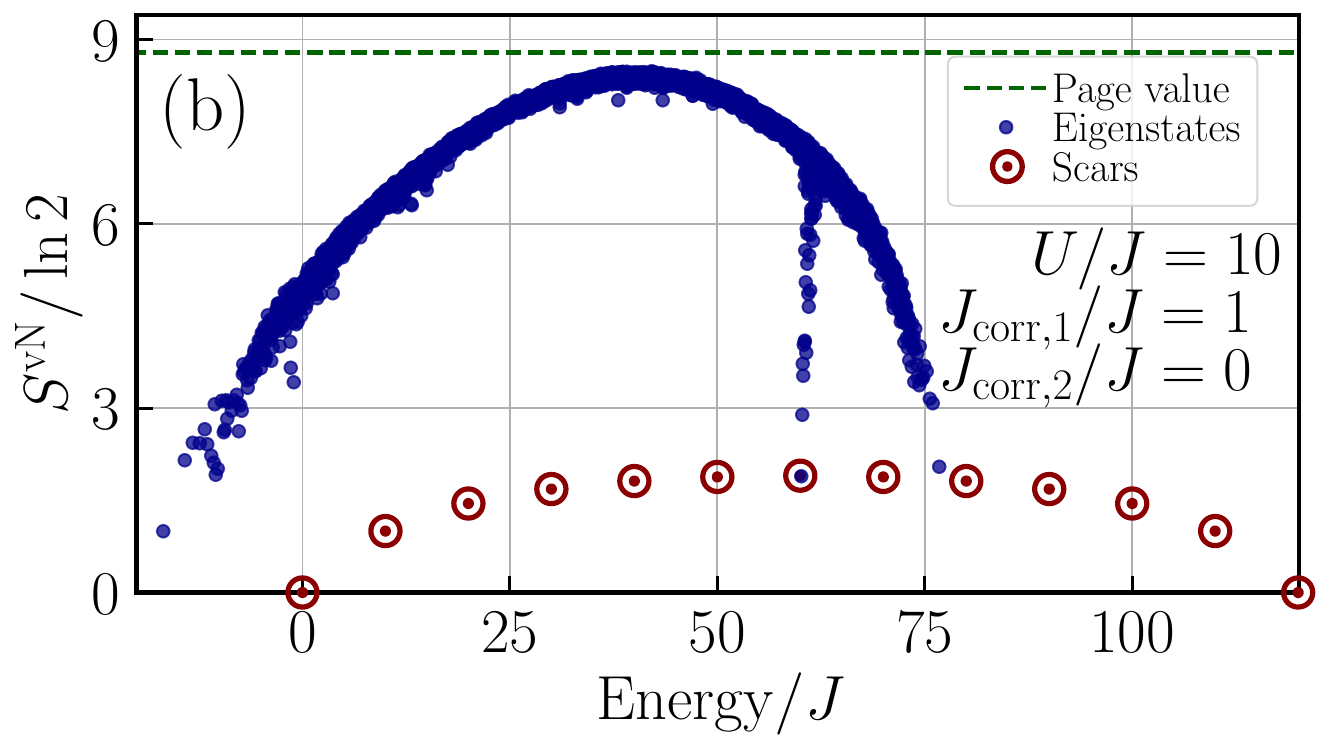}
\\
\includegraphics[width=\columnwidth]{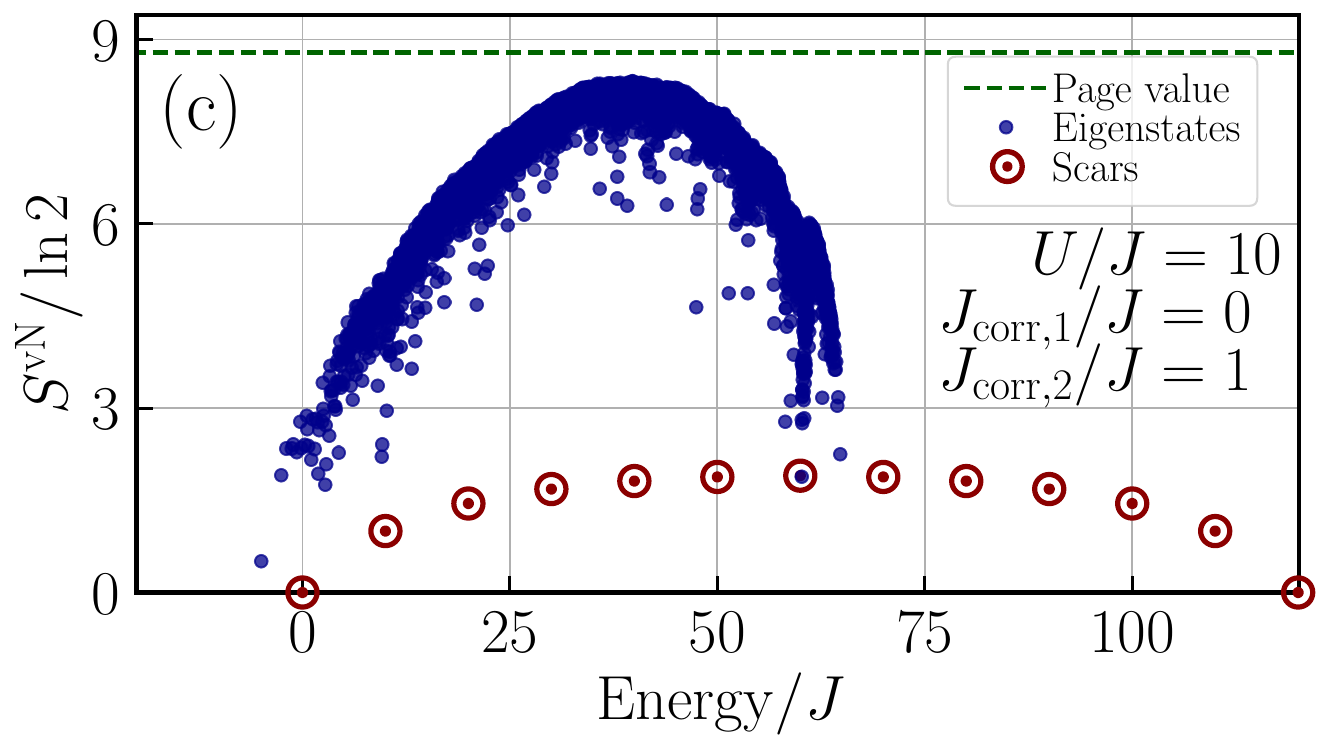}
\caption{%
EE as a function of energy
in the presence of correlated hoppings.
We consider the system size $L=12$
under open boundary conditions
and the interaction strength $U/J=10$
in Eq.~\eqref{supp:eq:generalized_bose_hubbard_model}.
The correlated hopping is chosen as
(a) $(J_{\rm corr,1}/J,J_{\rm corr,2}/J)=(0,0)$,
(b) $(J_{\rm corr,1}/J,J_{\rm corr,2}/J)=(1,0)$,
and
(c) $(J_{\rm corr,1}/J,J_{\rm corr,2}/J)=(0,1)$
[see Fig.~\ref{supp:fig:points}].
The quantum number sector with the particle number
$N=L$ (unit filling)
and the even parity $\mathcal{I}=+1$
is given by blue dots.
The largest EE almost saturates at the Page
value~\cite{page1993}
$S^{\rm Page} = (\ln 3)L/2 - 1/2$~\cite{schecter2019,chattopadhyay2020}
(a green dashed line).
Each QMBS state (a red circle with a dot)
with the particle number $N$
has the energy $UN/2$
and the expectation values
$\langle\sum_{i=1}^{L} \hat{n}_i\rangle = N$ and
$\langle\sum_{i=1}^{L} \hat{n}_i^2\rangle = 2N$.
Its half-chain
EE at $N=L$, corresponding to a state $|S_{n=L/2}\rangle$,
becomes $S^{\rm vN}_{A}\rightarrow [\ln(\pi L/8)+1]/2$
for $L\rightarrow\infty$.
Irrespective of the choice of the correlated hopping parameters
$J_{\rm corr,1}$ and $J_{\rm corr,2}$,
each QMBS state has the same energy and the same
EE.}
\label{supp:fig:scars_corr}
\end{figure}

To this end,
we calculate all the eigenenergies
for a certain symmetry sector of the models
by the exact diagonalization method.
To carefully capture the possible integrable-like behavior
in the energy spectrum, we choose the even total number sector,
corresponding to the even total magnetization sector
in the $S=1$ $XY$ model.
Hereafter, we display the results for
the particle number $N=L+2$
in the even parity sector ($\mathcal{I}=+1$)
under open boundary conditions.
The system size is chosen to be $L=10$ and $L=12$,
whose dimensions of the Hilbert spaces are
$3405$ and $29202$, respectively.

We show the distribution $P(r)$ at $L=12$
for the models
$\hat{H}^{\rm non}(\alpha=1)$,
$\hat{H}^{\rm non}(\alpha=0)$,
$\hat{H}^{\rm non,1}$, and
$\hat{H}^{\rm non,2}$
in Fig.~\ref{supp:fig:distribution_pr}.
In the case of $\hat{H}^{\rm non}(\alpha=1)$,
which also corresponds to
the $S=1$ $XY$ model having the integrable-like behavior
in even magnetic sectors,
$P(r)$ almost obeys the Poisson distribution.
The small deviation from the expected Poisson distribution
may be due to the presence of unresolved symmetries
in even total number sectors~\cite{chattopadhyay2020}.
Therefore, it is plausible that
the model $\hat{H}^{\rm non}(\alpha=1)$ is integrable-like.
On the other hand,
$P(r)$'s of the other models
obey the Wigner-Dyson distribution,
suggesting that these systems are nonintegrable.

We also calculate the mean level spacing ratio $r_{\rm mean}$
for the model $\hat{H}^{\rm non}(\alpha)$
in Fig.~\ref{supp:fig:r_mean}.
At $\alpha=1$, the model is equivalent to
the $S=1$ $XY$ model.
As we expected, the integrable-like behavior also appears
in the mean level spacing,
and $r_{\rm mean}$ takes the value
close to that of the Poisson distribution
($r_{\rm mean}\approx 0.386$).
On the other hand,
for $\alpha<1$,
$r_{\rm mean}$ is nearly equal to
that of the Wigner-Dyson distribution
($r_{\rm mean}\approx 0.536$).
These results indicate that
$\hat{H}^{\rm non}(\alpha)$ is nonintegrable
except for the special point $\alpha=1$.

\subsection{Bipartite EE}
\label{supp:sec:ent_ent}

We numerically confirm the existence of
the QMBS states
satisfying the relations predicted in Sec.~\ref{supp:sec:athermal_eigenstates}.
On some of the points that we have chosen
[see Fig.~\ref{supp:fig:points}],
we show the energy dependence of the
EE for the Hamiltonian
in Eq.~\eqref{supp:eq:generalized_bose_hubbard_model}
with $V=\mu=0$
[see Fig.~\ref{supp:fig:scars_corr}
and Fig.~1 in the main text].
All the eigenstates in the even number of total particles
($N=0$, $2$, $4$, \dots, $2L-2$, $2L$) are calculated
for the model with the open boundary condition
by the exact diagonalization method
using the \textsc{QuSpin} library~\cite{weinberg2017,weinberg2019}.
In particular,
we calculate the
even parity $\mathcal{I}=+1$
(odd parity $\mathcal{I}=-1$)
sector
when $N=4m$ ($N=4m+2$)
with $m$ being an integer.
The EE is obtained
from the singular value decomposition
of a wave function utilizing $\rm U(1)$
symmetry~\cite{jung2020},
associated with the conservation of the total particle
number~\cite{schnack2008,zhang2010,szabados2012,raventos2017}.
We specifically display all the eigenstates with $N=L$
and the QMBS states
for $N=0$, $2$, $4$, \dots, $2L-2$, $2L$.

For a weaker interaction ($U/J=1$)
in the case of the Hamiltonian $H$ in Eq.~\eqref{supp:eq:hubbard_model_h},
the data points for different $N$ distribute similarly
[see Fig.~1 in the main text].
In each sector $N$,
the QMBS state
with the energy $UN/2$
exhibits the area-law
EE with a logarithmic correction.
On the other hand, for a stronger interaction ($U/J=10$)
[see Fig.~\ref{supp:fig:scars_corr}(a)],
we observe a more separated structure
(EE peaks around the energy $\approx 10J$, $20J$, $30J$, and $40J$)
in the energy spectra
than in the case of $U/J=1$.
Even in this case, we still confirm the presence of the QMBS states.

We also numerically examine the effect of the correlated hopping terms.
We calculate all eigenstates of the generalized Bose-Hubbard model
in Eq.~\eqref{supp:eq:generalized_bose_hubbard_model}
with $V=\mu=0$.
We consider the system with $L=12$ at $U/J=10$
as in the case of Fig.~\ref{supp:fig:scars_corr}(a).
We choose only the correlated hopping
$J_{\rm corr,1}/J=1$ in Fig.~\ref{supp:fig:scars_corr}(b),
whereas
only
$J_{\rm corr,2}/J=1$ in Fig.~\ref{supp:fig:scars_corr}(c),
in addition to the conventional hopping
$J$.
They are nonintegrable
as we have confirmed in
Sec.~\ref{supp:sec:nonintegrability}.
These spectra are less separated than in the case of
Fig.~\ref{supp:fig:scars_corr}(a)
because the additional correlated hopping terms
weaken the effect of interaction.
The athermal eigenstates still remain 
in the presence of the correlated hopping term
and have exactly the same energy
as that without the correlated hoppings.

\begin{figure}[!ht]
\centering
\includegraphics[width=\columnwidth]{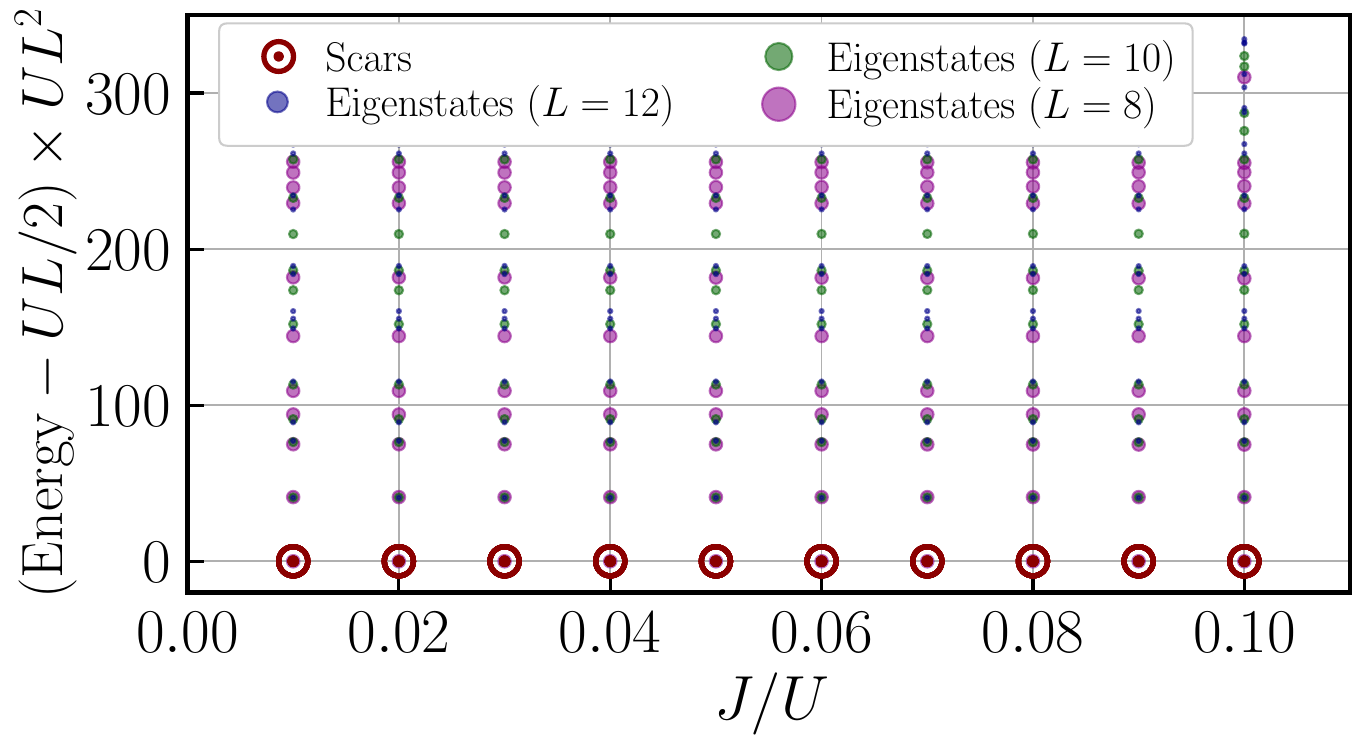}
\caption{%
Magnified energy spectra as a function of
the interaction strength at unit filling.
The parameters are the same as in Fig.~2 of the main text.
}
\label{supp:fig:ene_spectra_magnified}
\end{figure}

\subsection{Size dependence of the energy spectra}
\label{supp:sec:size_dep}

We check the finite-size effect of the energy spectra 
in the strong-coupling limit.
We focus on the system sizes up to $L=12$.
At unit filling for the symmetry sector where the QMBS state exists,
the dimension of the Hilbert space is $563$ for $L=8$,
$4451$ for $L=10$,
and $36965$ for $L=12$.

We multiply the energy by $UL^2$
to inspect the excitations clearly,
as shown in Fig.~\ref{supp:fig:ene_spectra_magnified}.
In the strong-coupling limit,
the lower excitations are essentially characterized by
those of the ferromagnetic Heisenberg model with the spin exchange
interaction, which is proportional to $J^2/U$ (see the main text).
Since the excitation of the ferromagnetic Heisenberg model on a lattice
grows as $k^2$ for momenta $k=2n\pi/L$ with a small integer number $n$,
we expect that the lowest excitation in a finite system
scales as $1/(UL^2)$ in the unit of $J$.
This behavior is correctly observed,
and the scaled lowest excitations show very small size and interaction
dependencies for $L\ge 8$ and $J/U\le 0.1$.

\subsection{Effect of external potential}
\label{supp:sec:effect_ext_pot}

\begin{figure}[!h]
\centering
\includegraphics[width=\columnwidth]{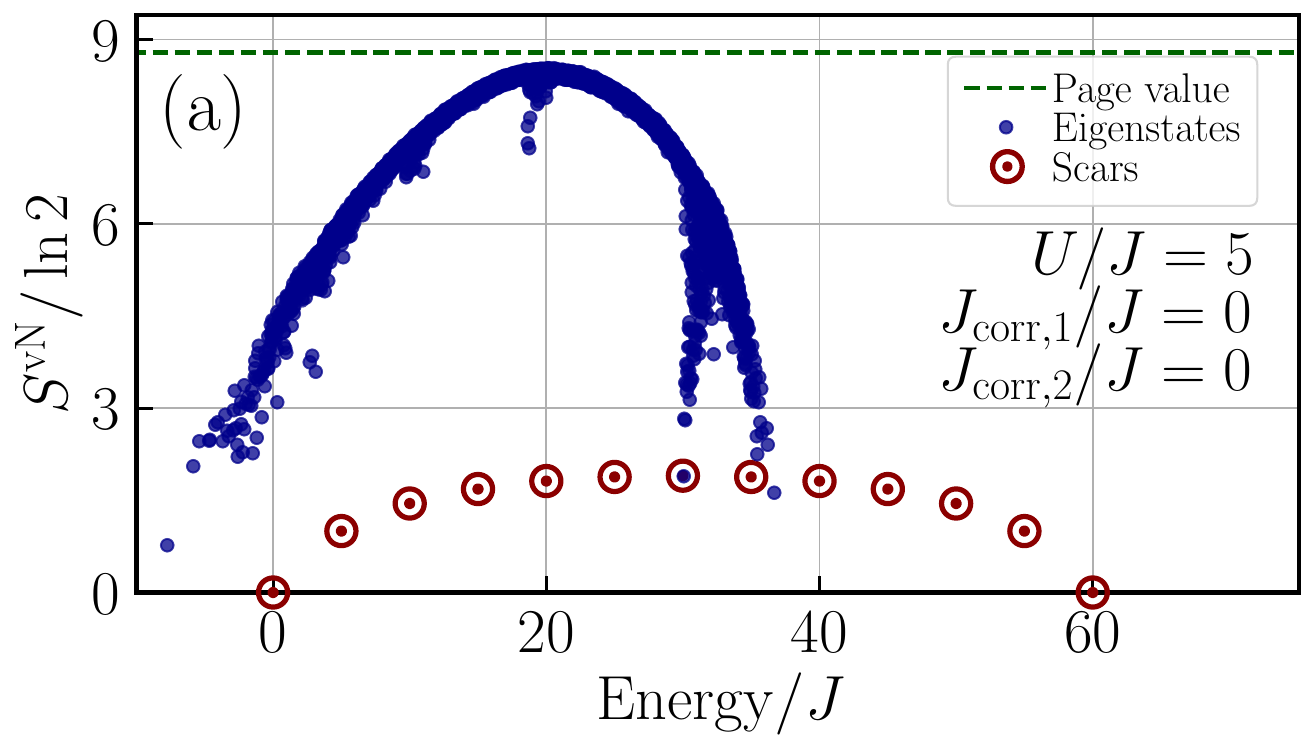}
\hfil  
\includegraphics[width=\columnwidth]{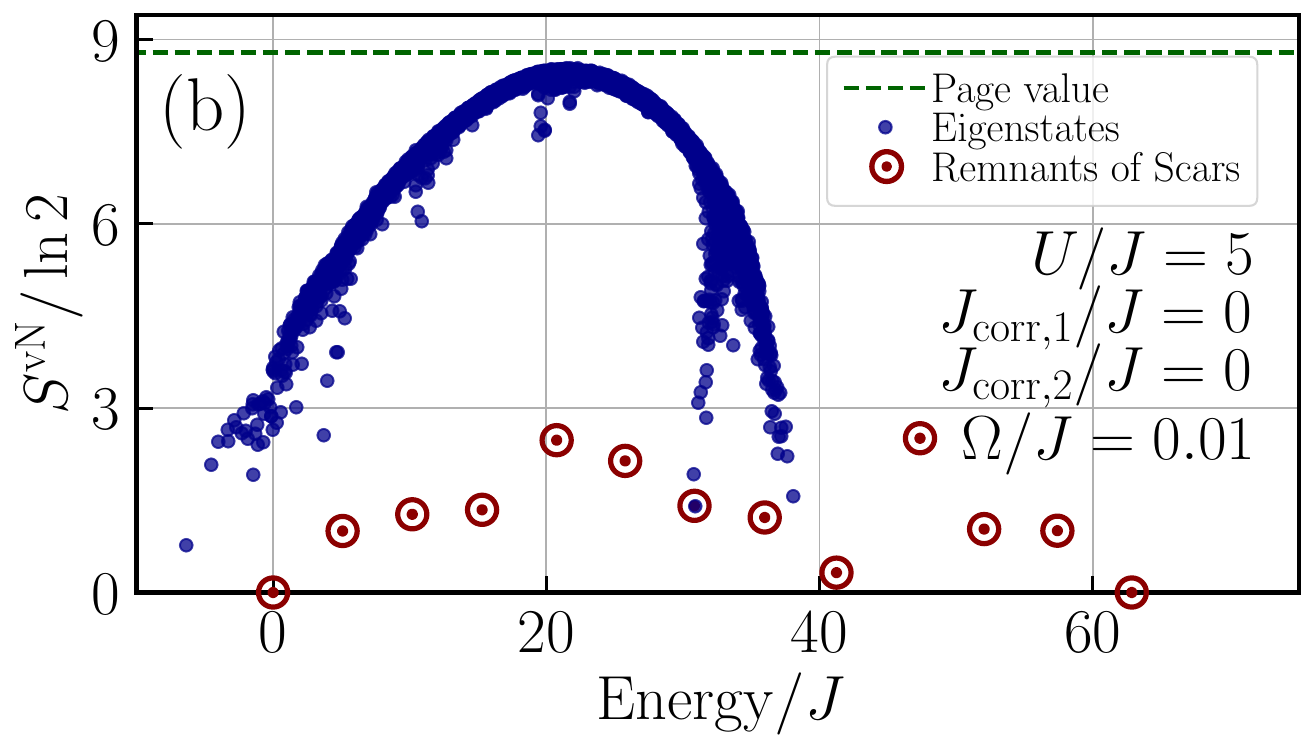}
\caption{%
Entanglement entropy as a function of energy for
(a) without external potential
and
(b) with external potential.
We consider the system size $L=12$
under open boundary conditions
and the interaction strength $U/J=5$
in Eq.~\eqref{supp:eq:generalized_bose_hubbard_model}.
Remnants of scars are characterized by
the largest $\langle \sum_{i=1}^{L} \hat{n}_i^2\rangle (\approx 2N)$
in each sector of particle number
$\langle \sum_{i=1}^{L} \hat{n}_i\rangle = N$.
}
\label{supp:fig:ext_pot_1}
\end{figure}

\begin{figure}[!h]
\centering
\includegraphics[width=\columnwidth]{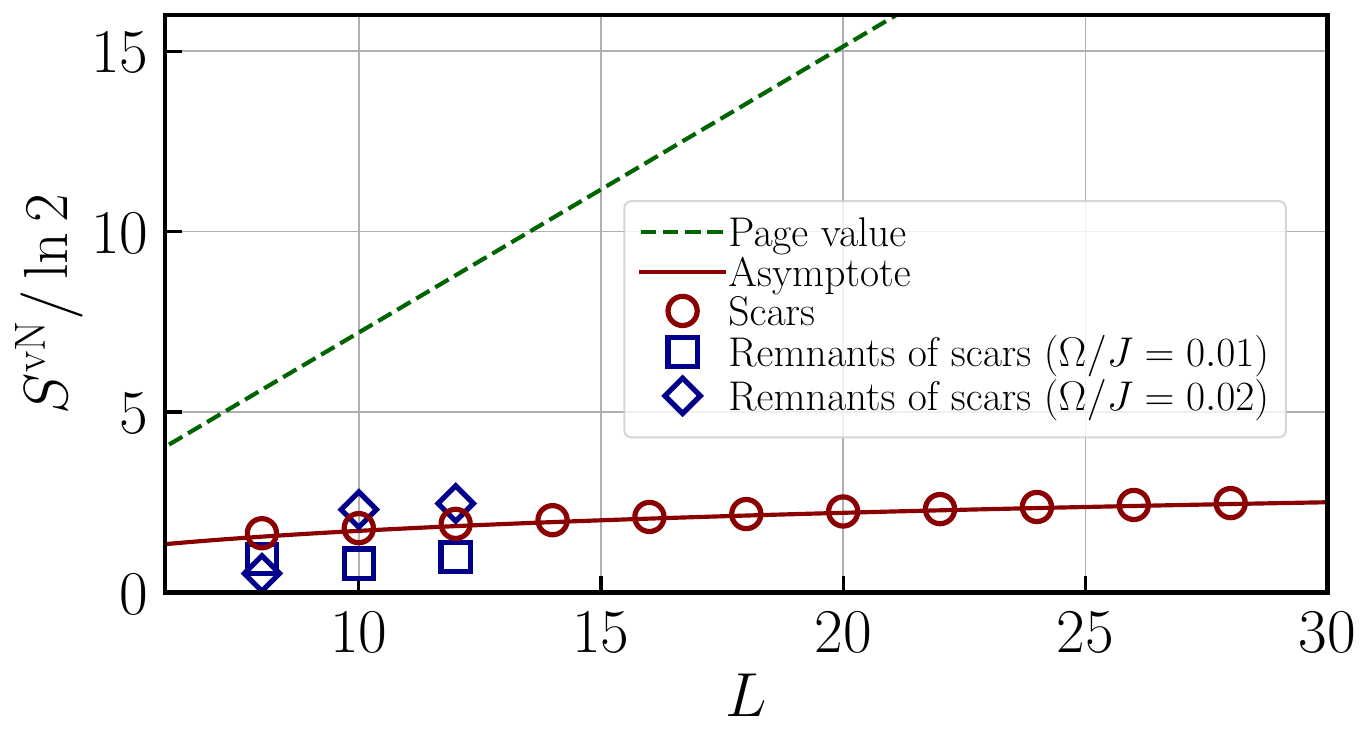}
\caption{%
Size dependence of entanglement entropy of scars
with and without external potential.
The entanglement entropy grows in a similar manner
regardless of the presence or absence of the external potential.
}
\label{supp:fig:ext_pot_2}
\end{figure}

We examine the half-chain entanglement entropy
in the presence of the parabolic potential.
We consider the Hamiltonian
\begin{align}
\label{supp:eq:hamiltonian_omega}
 \hat{H} =
 - J \sum_{i=1}^{L-1} ( \hat{a}^{\dagger}_i \hat{a}_{i+1}
 + \hat{a}^{\dagger}_{i+1} \hat{a}_i )
 + \sum_{i=1}^{L} \Omega_i \hat{n}_i
 + \frac{U}{2} \sum_{i=1}^{L} \hat{n}_i (\hat{n}_i - 1)
\end{align}
and choose the external potential
$\Omega_i = \Omega [i-(L+1)/2]^2$,
where $\Omega>0$ is the strength of the potential.
In the absence of the potential ($\Omega=0$),
at each filling ($\langle \sum_{i=1}^{L} \hat{n}_i\rangle = N$),
we numerically find that the QMBS state satisfies
$\langle \sum_{i=1}^{L} \hat{n}_i^2\rangle = 2N$,
whereas other eigenstates are found to exhibit
$\langle \sum_{i=1}^{L} \hat{n}_i^2\rangle < 2N$.
This observation suggests that,
in each particle number sector,
the eigenstate having the largest
$\langle \sum_{i=1}^{L} \hat{n}_i^2\rangle$ corresponds to the QMBS
state.
In the presence of small potential ($0<\Omega\ll J$),
we anticipate this trend to continue
and can safely regard
the eigenstate with the largest
$\langle \sum_{i=1}^{L} \hat{n}_i^2\rangle$
as the remnant of the QMBS state.
Indeed, such state still exhibits a very small entanglement entropy.
As shown in Fig.~\ref{supp:fig:ext_pot_1},
in the presence of a small external potential ($\Omega/J=0.01$),
we find that remnants of the QMBS states exhibit
entanglement entropies much smaller than those for volume-law states.

We also investigate the size dependence of
the half-chain entanglement entropy at unit filling
in the presence of the parabolic potential
(see Fig.~\ref{supp:fig:ext_pot_2}).
The exact QMBS state without the potential shows the entanglement
entropy
$S^{\rm vN}_{A} = - \sum_{k=0}^{L/2} \lambda_k \ln \lambda_k$
with $\lambda_k = \binom{L/2}{k}\binom{L/2}{L/2-k}/\binom{L}{L/2}$,
which nearly grows as
$S^{\rm vN}_{A} \approx [\ln(\pi L/8)+1]/2$ for a sufficiently large
size $L$~\cite{schecter2019}.
This value is much smaller than the Page value~\cite{page1993}
$S^{\rm Page} = (\ln 3)L/2 - 1/2$~\cite{schecter2019,chattopadhyay2020}.
In a similar manner, in the presence of the external potential
($\Omega/J=0.01$, $0.02$),
the remnants of the QMBS states show the entanglement entropy
close to the one with the QMBS state without the potential.
Within the sizes that we could study,
they seem to be consistent with the logarithmic size dependence
of the entanglement entropy growth.
The presence of small external potential does not affect the QMBS states
significantly.

Note that
the Bose-Hubbard simulator in a flat box potential
has been realized very recently~\cite{impertro2023,wienand2023_arxiv}.
Therefore, one may prepare the ideal QMBS states
without any potential effects experimentally.

\onecolumngrid